\documentclass[aps,prl,reprint]{revtex4-2}
\usepackage{blindtext}
\usepackage{graphicx}
\usepackage{subcaption}
\usepackage{xcolor}
\usepackage[fleqn]{amsmath}
\usepackage[font=small,skip=3pt]{caption}
\usepackage{makecell}
\usepackage{hyperref} 
\usepackage{cleveref}
\usepackage{url}
\begin{document}
\title{Prediction of ELM-free Operation in Spherical Tokamaks With High Plasma Squareness}
\author{J. F. Parisi$^{1}$}
\email{jparisi@pppl.gov}
\author{J. W. Berkery$^1$}
\author{K. Imada$^2$}
\author{A. O. Nelson$^3$} 
\author{S. M. Kaye$^1$}
\author{P. B. Snyder$^4$}
\author{M. Lampert$^1$}
\author{A. Kleiner$^1$} 
\affiliation{$^1$Princeton Plasma Physics Laboratory, Princeton University, Princeton, NJ, USA}
\affiliation{$^2$York Plasma Institute, Department of Physics, University of York, Heslington, York, United Kingdom}
\affiliation{$^3$Department of Applied Physics and Applied Mathematics, Columbia University, New York, NY, USA}
\affiliation{$^4$Oak Ridge National Laboratory, Oak Ridge, TN, USA}
\begin{abstract}
We predict that high plasma squareness in spherical tokamaks (STs) could result in edge-localized-mode (ELM)-free H-mode. The effect of squareness on gyrokinetic and peeling-ballooning-mode width-height pedestal scalings is calculated for STs. Because STs can sustain H-mode in first ballooning stability, first-stable pedestals with lower gradients may be further from the peeling-ballooning-mode boundary and therefore naturally free of Type 1 ELMs. We show that while higher squareness destabilizes ballooning modes in first stability, the ELM stability boundary is essentially unchanged. Therefore, higher squareness could result in ELM-free discharges. Random Forest (RF) machine learning models for the gyrokinetic growth rate and distance from first stability are used to predict how squareness affects stability. A RF model with only three easily obtainable geometric inputs predicts proximity to the gyrokinetic width-height scaling on a test dataset with high accuracy, $R^2 = 0.965$.
\end{abstract}

\maketitle

\setlength{\parskip}{0mm}
\setlength{\textfloatsep}{5pt}

\setlength{\belowdisplayskip}{6pt} \setlength{\belowdisplayshortskip}{6pt}
\setlength{\abovedisplayskip}{6pt} \setlength{\abovedisplayshortskip}{6pt}

\section{Introduction}

Tokamak power plants must balance competing design tradeoffs, one of the most critical being the conflict between achieving a high power density and ensuring a manageable power exhaust. The cost of electricity from a tokamak power plant is predicted to drop significantly with higher volume-averaged power density $\langle p_f \rangle_V$ \cite{Wade2021}. H-mode operation \cite{Wagner1982} offers a compelling route to increased $\langle p_f \rangle_V$ because the pressure pedestal near the plasma edge raises the overall core pressure disproportionately \cite{Snyder2019}. In contrast, in L-mode \cite{Austin2019,Nelson2023,Wilson2024} -- which has a reduced or absent edge pedestal -- core confinement must be enhanced to achieve similar power densities.

A major obstacle to using H-mode in future power plants is the presence of Type-1 edge-localized-modes (ELMs). These events, while manageable in present-day experiments, produce transient heat fluxes that are unacceptable in power plants \cite{Federici2019b,Creely2020,Muldrew2024,Maingi_2014,Hughes2020,Kuang2020,Viezzer2023}. Therefore, ELM-free or small-ELM H-modes are a high priority to avoid the prohibitively large heat loads associated with Type-1 ELMs.

There are several candidates for power plant-relevant ELM-free or small-ELM regimes, each with its own set of challenges \cite{Viezzer2023}. The quasi-continuous exhaust (QCE) regime on ASDEX-U features ballooning instability near the separatrix \cite{Faitsch2021,Harrer_2022,Radovanovic_2022,Faitsch2023,Dunne2024}. Enhanced-pedestal H-mode was discovered on NSTX following lithium conditioning \cite{Maingi2009,Maingi2010,Canik2013,Gerhardt_2014}, leading to wide-pedestal ELM-free regimes. Enhanced D-alpha (EDA) H-mode is another ELM-free regime believed to be limited by a quasi-coherent mode \cite{Hubbard2001,LaBombard_2014,Gil_2020,Macwan2024}, whereas I-mode features H-mode-like temperature confinement but L-mode-like particle confinement \cite{Greenwald1998,Ryter1998,Whyte2010,Hubbard_2017}. Quiescent H-mode \cite{Burrell2001,Sakamoto_2004, Suttrop_2005, Garofalo_2011} and its variants can also operate without ELMs, relying on edge harmonic oscillations or turbulence-limited mechanisms \cite{Ernst2024,Burrell2016,Chen2017b,Wilks2021b,Houshmandyar2022}. In parallel, active ELM suppression strategies such as resonant magnetic perturbations \cite{Kirk_2010}, vertical kicks \cite{Degeling_2003}, ELM pacing \cite{Evans_2004}, and supersonic molecular beam injection \cite{LianghuaYao2001} -- continue to be pursued. Finally, L-mode remains an ELM-free regime of interest but typically features lower core power density \cite{Jardin2000,Kikuchi2019,Austin2019,Nelson2023,Paz-Soldan_2024,Wilson2024}. Although impurity flushing is a major challenge for ELM-free regimes \cite{Putterich2011}, we do not address that here.

Additionally, pedestal performance cannot be considered in isolation from the plasma boundary and divertor region \cite{Snyder2024,Zhang_2024,Chang_2024}. Even ELM-free regimes will have high steady state power fluxes and neutron fluences, requiring integrated solutions for both core confinement and power exhaust. Advanced divertor geometries (e.g., snowflake, X-divertor, super-X \cite{Ryutov_2007,Kotschenreuther_2010,Kotschenreuther_2013,Kool2024}), improved fueling schemes \cite{Lang_2004,Field_2021}, and the use of liquid metal walls \cite{Morley_2004,Horacek_2021,Maingi2025} are among the promising strategies for handling the intense, steady-state heat loads in fusion power systems.

Accurate prediction of ELM-free pedestal behavior remains an open problem, particularly given that the fusion gain $Q$ in devices such as SPARC is highly sensitive to pedestal-top conditions \cite{Snyder2019,Creely2020,Hughes2020}. Operational uncertainty in pedestal modeling can thus pose a significant risk to performance. Existing models suggest that pedestal profiles can be limited by MHD stability \cite{Snyder2009,Parisi_2024}—assuming stiff transport—or by transport processes, where gradients are set by plasma sources and instabilities \cite{kotschenreuther2024transport}. For ELMy discharges, the EPED model \cite{Snyder2002,Snyder2009,Snyder2011,Beurskens2011,Walk2012,Leyland2013} predicts the ELM-cycle-maximum pedestal pressure width $\Delta_{\mathrm{ped}}$ and height $\beta_{\theta,\mathrm{ped}}$ using two stability thresholds: nearly local kinetic-ballooning-modes (KBMs) \cite{Connor1979,Tang1980,Hastie1981,Dong1999,Pueschel2008,Snyder2009,Coury2016,Ma2017,Aleynikova2018}, which limit the pressure gradient on a given flux surface and peeling-ballooning modes (PBMs) \cite{Lortz1975,Connor1998,Wilson1999}, which limit the current and pressure gradient profiles across the pedestal. Crossing the PBM stability limit is thought to trigger Type-1 ELMs.

Further complicating matters, pedestal transport and sources are strongly coupled: for instance, increased particle transport into the scrape-off layer (SOL) can raise the pedestal particle source yet concurrently siphon away energy for ionization \cite{Snyder2024}. Recent predictive models for QCE and EDA H-mode access and separatrix conditions are promising \cite{Eich_2021,Eich2024arxiv}, with recent validation on Alcator C-Mod \cite{Miller2024arxiv}.

\begin{figure}[tb]
    \centering
    \includegraphics[width=0.37\textwidth]{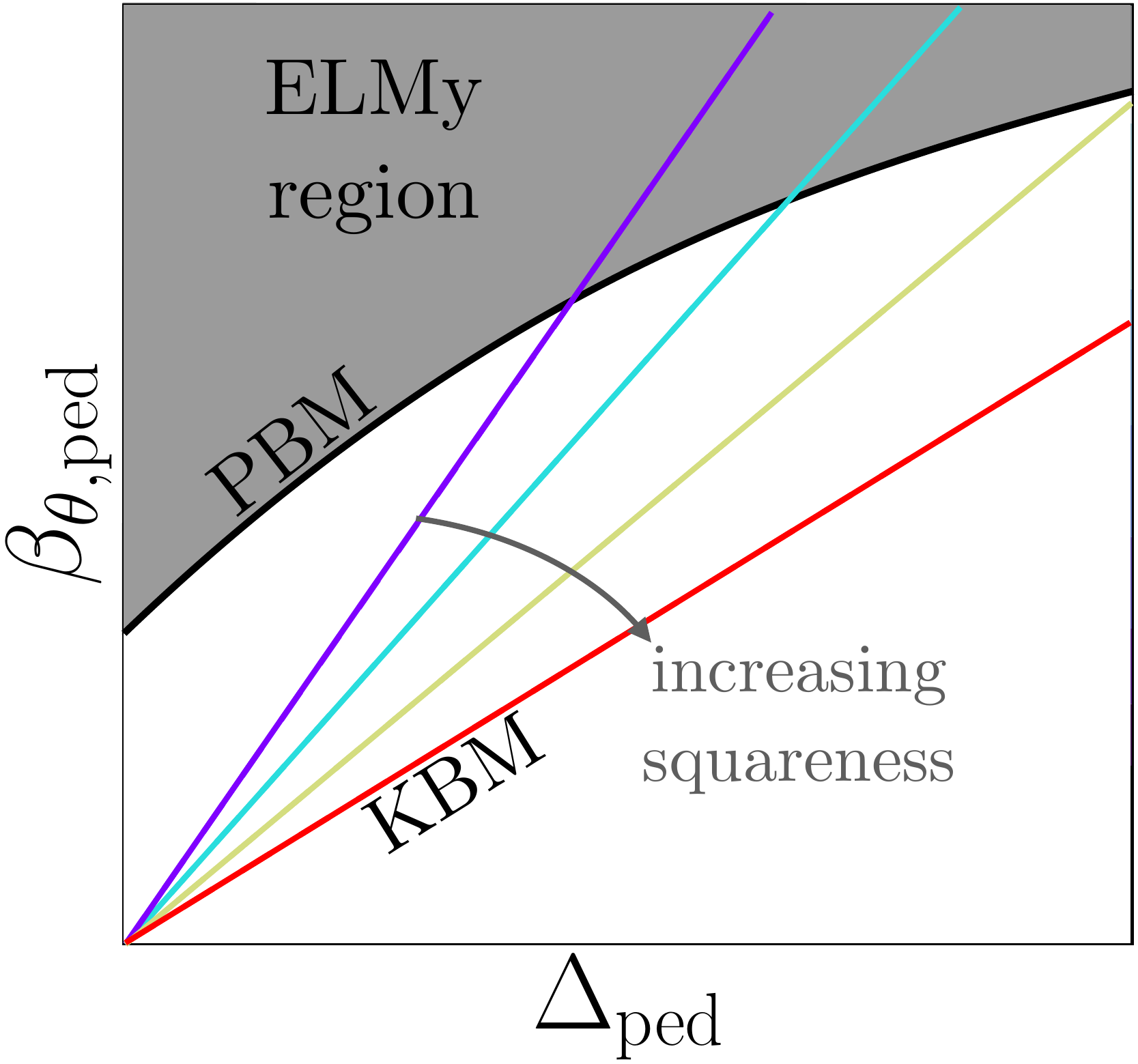}
    \caption{Schematic effect of squareness on PBM (solid) and KBM (dashed) stability boundary in pedestal width ($\Delta_{\mathrm{ped} }$) and height ($\beta_{\theta,\mathrm{ped} }$) space. The PBM boundary is relatively invariant to changes in squareness in MAST-U 48339.}
    \label{fig:schematic_squareness}
\end{figure}

Plasma shaping and aspect ratio \cite{Parisi_2024b,Snyder2024}, known to play major roles in the core, pedestal, and SOL \cite{Goldston1984,Kaye1985,Waltz1999,Kinsey_2007,Belli2008,Angelino2009,Marinoni2009,Jolliet2014,Ball2015,Snyder2015,Merle2017,Austin2019,Riva2020,Nelson2023,Nelson2024b,Nelson2024c,Mariani_2024,Balestri_2024,Sun_2024}, may offer a tool to avoid the ELM limit. Because spherical tokamaks (STs) \cite{Peng1986,Harrison_2019,Doyle_2021,Sontag_2022,Berkery_2024,Chapman_2024,McNamara_2024} can sustain H-mode in first stability \cite{Parisi_2024,Nelson2024}, they may achieve high pedestal pressure without encountering Type-1 ELMs. Somewhat counterintuitively, a first-stable but lower-gradient pedestal might attain comparable or even higher total pedestal pressure, whereas a narrow pedestal with steeper gradients could saturate at the ELM stability limit. While such a strategy may not be universal across all aspect ratios, it offers an intriguing path toward H-mode operation in STs.

In this paper, we demonstrate that shaping could be a major lever for power-plant–relevant H-mode operation. We show that at low aspect ratio, increasing plasma squareness moves the pedestal away from the ELMy regime by modifying the KBM stability boundary while leaving the PBM boundary nearly unchanged. This idea is an extension of those presented in \cite{Nelson2022,Parisi_2024b}, and has a different result to \cite{Snyder2015,Merle2017} where plasma triangularity had a strong effect on the PBM boundary. \Cref{fig:schematic_squareness} illustrates how higher squareness lowers the KBM stability threshold without substantially shifting the PBM boundary. Readers seeking a quick summary can refer to \Cref{fig:squareness_first_second_stab_MASTU}, where we substantiate the schematic in \Cref{fig:schematic_squareness} with simulation data, demonstrating how shaping can lead to inherently ELM-free H-mode operation in low-aspect-ratio tokamaks. This idea has already been explored experimentally for triangularity on MAST-U \cite{Nelson2024c}. In this work, we focus on squareness.

Ultimately, an integrated approach—combining shaping, fueling, and power mitigation techniques—will be necessary to ensure robust performance with manageable power exhaust. By focusing on the connection between pedestal stability and plasma shaping, we aim to provide a potential path forward for H-mode in future STs with high plasma squareness.

The layout of this work is as follows. We give an overview of plasma squareness in \Cref{sec:squareness} and gyrokinetics \Cref{sec:GK_intro}. We present the effect of squareness on the pedestal gyrokinetic width-height scaling in \Cref{sec:GKwidthheight}. We then find the combined pedestal gyrokinetic and peeling-ballooning width-height scalings in \Cref{sec:GK_peeling_widthheight}. In \Cref{sec:random_tree_model}, we use Random Forest models to predict the effect of geometry on gyrokinetic growth rates and proximity to the gyrokinetic width-height scaling. We summarize in \Cref{sec:summary}. In \Cref{sec:peddefinitions} we give detailed definitions of pedestal parameters, in \Cref{sec:geodefinitions} we describe the normalization of gyrokinetic quantities, and in \Cref{sec:eigen_average} we show results from a Random Forest model for the KBM linear growth rate that uses eigenmode-averaged geometric coefficients as input features.

\section{Plasma Squareness} \label{sec:squareness}

\begin{figure}[!tb]
    \centering
    \begin{subfigure}[t]{0.23\textwidth}
    \centering
    \includegraphics[width=1.0\textwidth]{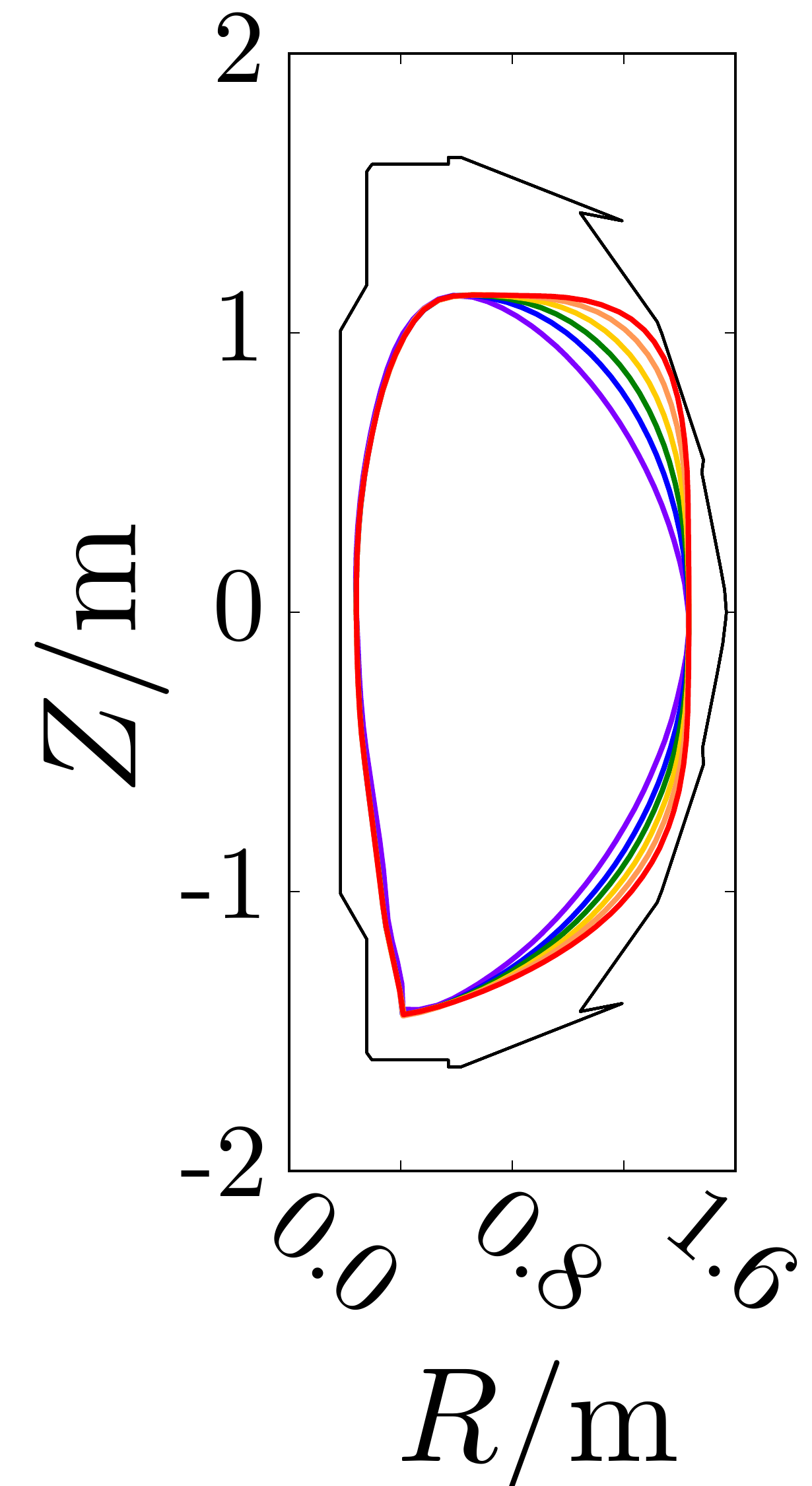}
    \caption{NSTX 132543}
    \end{subfigure}
    \begin{subfigure}[t]{0.23\textwidth}
    \centering
    \includegraphics[width=1.0\textwidth]{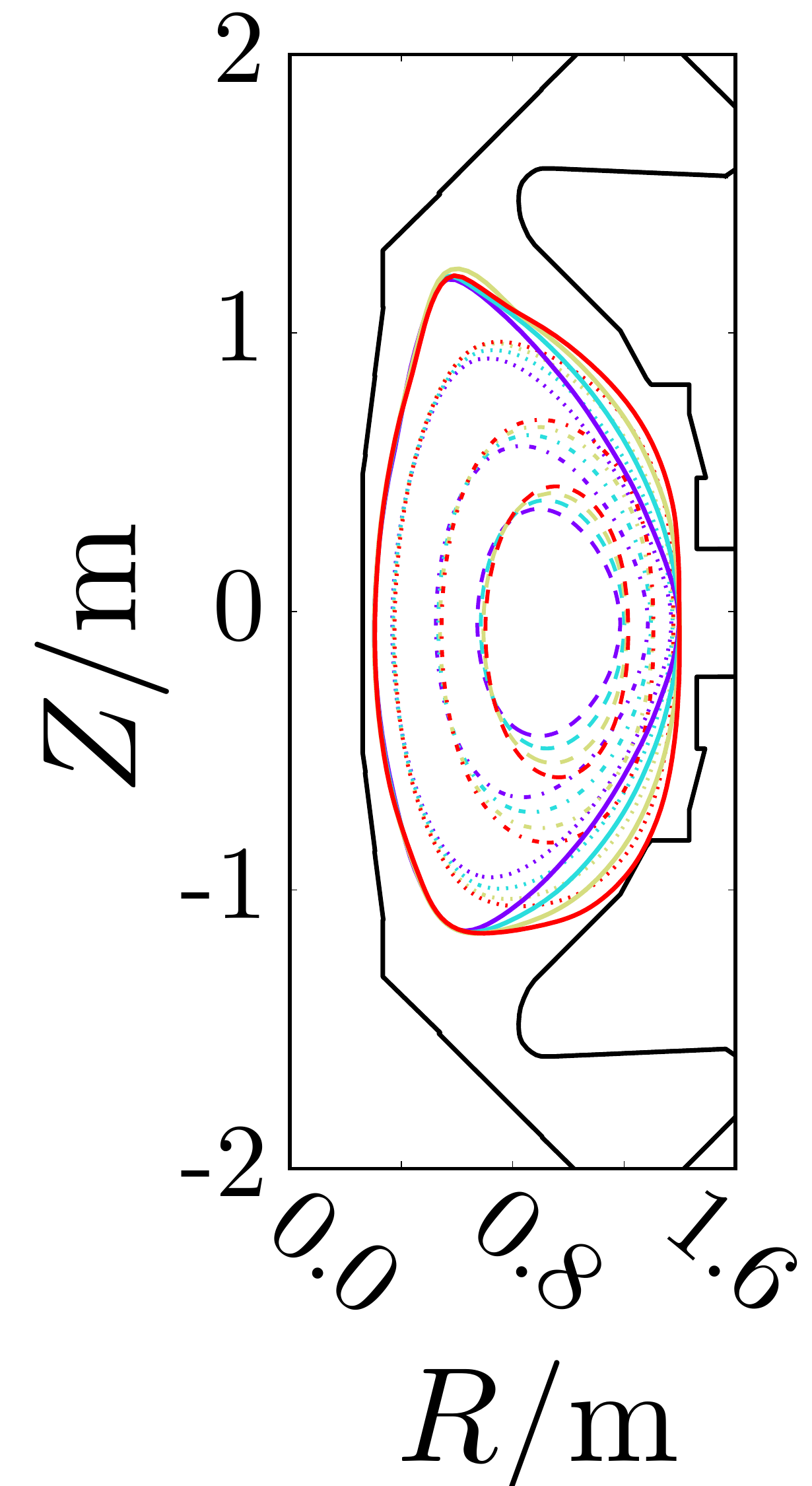}
    \caption{MAST-U 48339}
    \end{subfigure}
     ~
    \caption{Scan in plasma outer squareness for (a) NSTX 132543 and (b) MAST-U 48339 (b). In (b), the flux surfaces $\psi_N = [0.3, 0.6, 0.9, 1.0]$ are plotted.}
    \label{fig:squarenesseq}
\end{figure}

\begin{figure}[!tb]
    \centering
    \begin{subfigure}[t]{0.23\textwidth}
    \centering
    \includegraphics[width=1.0\textwidth]{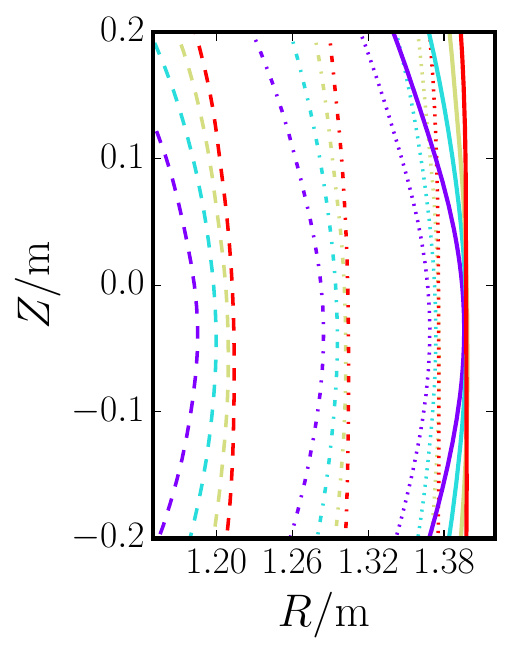}
    \caption{}
    \end{subfigure}
    \begin{subfigure}[t]{0.23\textwidth}
    \centering
    \includegraphics[width=1.0\textwidth]{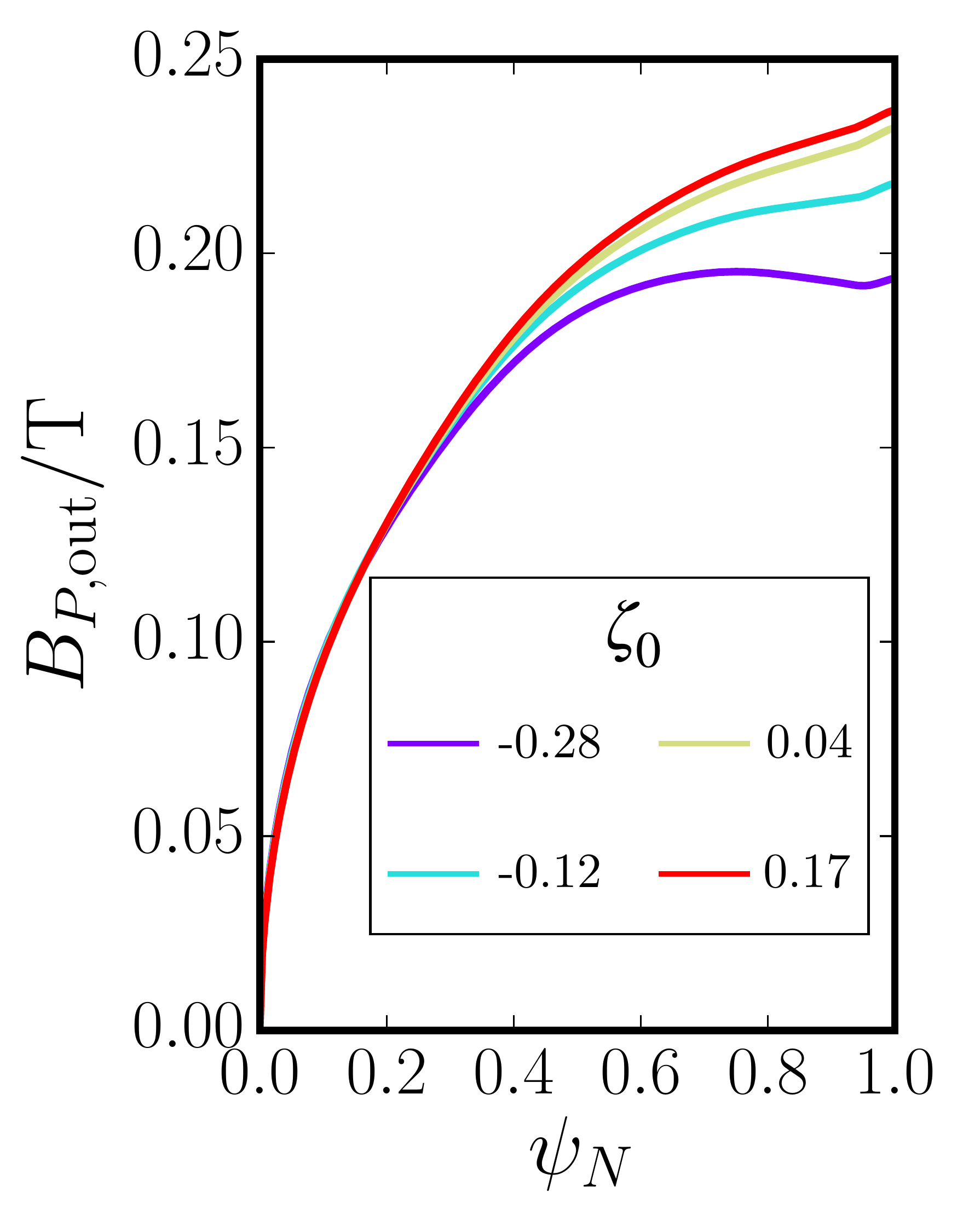}
    \caption{}
    \end{subfigure}
    \centering
    \begin{subfigure}[t]{0.23\textwidth}
    \centering
    \includegraphics[width=1.0\textwidth]{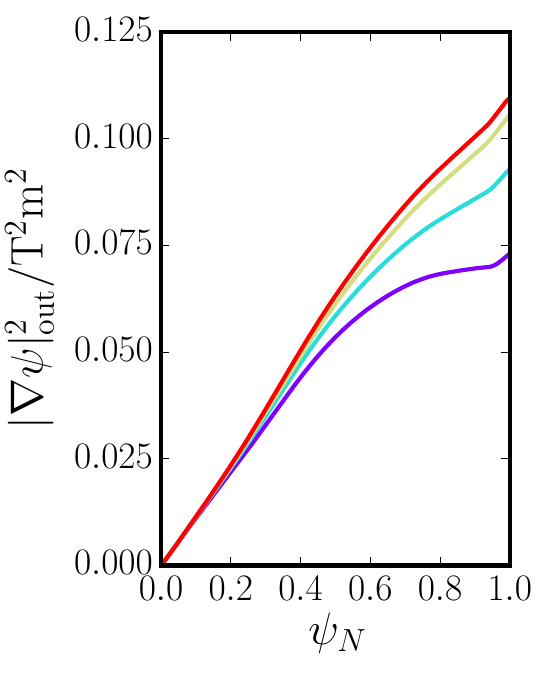}
    \caption{}
    \end{subfigure}
    \begin{subfigure}[t]{0.23\textwidth}
    \centering
    \includegraphics[width=0.99\textwidth]{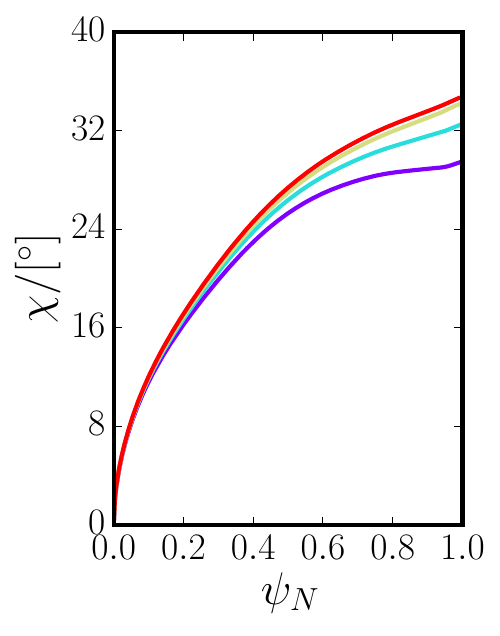}
    \caption{}
    \end{subfigure}
    \centering
    \begin{subfigure}[t]{0.23\textwidth}
    \centering
    \includegraphics[width=1.0\textwidth]{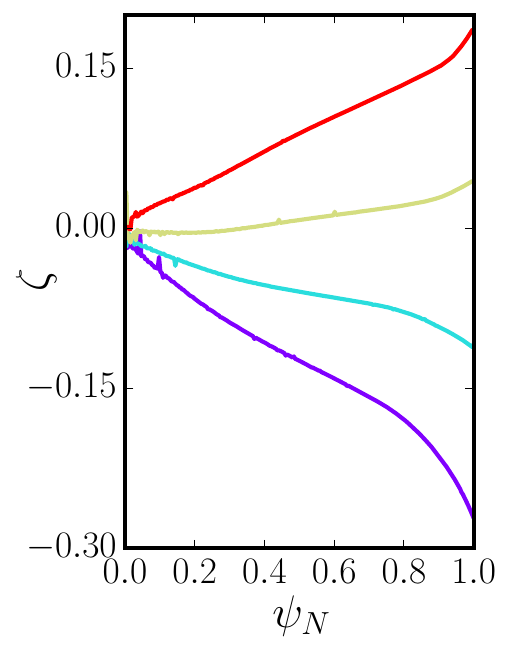}
    \caption{}
    \end{subfigure}
    \begin{subfigure}[t]{0.23\textwidth}
    \centering
    \includegraphics[width=1.0\textwidth]{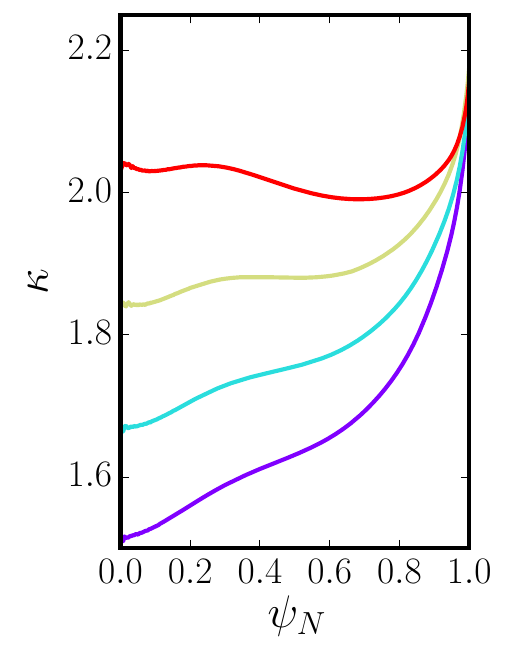}
    \caption{}
    \end{subfigure}
    \caption{Quantities for MAST-U 48339 with four different squareness values corresponding to \Cref{fig:squarenesseq}(b): (a) Flux surfaces zoomed in to a region near the outboard midplane. (b)-(f) show equilibrium quantities versus $\psi_N$: (b) Poloidal magnetic field at the outboard midplane (OMP), (c) OMP flux expansion, (d) OMP pitch angle $\chi$, (e) plasma squareness, (f) plasma elongation.}
    \label{fig:squarenesseq_effxMASTU}
\end{figure}

In this section, we give an overview of plasma squareness. For this work, we define the plasma squareness \cite{Turnbull1999} as
\begin{equation}
\zeta_0 = \arcsin \left( Z / \kappa_0 a \right) - \pi/4,
\end{equation}
where $Z$ is the vertical spatial coordinate, $\kappa_0$ is the plasma elongation, and $a$ is the minor radius. In \Cref{fig:squarenesseq}(a), we plot equilibria with six different squareness values, starting from an initial equilibrium corresponding to NSTX 132543 at 614ms and in \Cref{fig:squarenesseq}(b), equilibria with four different squareness values, starting from an initial equilibrium corresponding to MAST-U 48339 at 600ms. Both equilibria with nominal experimental squareness values ($\zeta_0 = -0.04$ for MAST-U and $\zeta_0 = 0.12$ for NSTX) are close to the ELM boundary. In this work, we focus on the outer plasma squareness, which is the poloidal mode number $m \approx 8$ shaping moment. Each equilibrium is reconstructed with a conserved total plasma current $I_p$ and normalize plasma beta, $\beta_N$. The total current is conserved by first recalculating the bootstrap current and rescaling the non-bootstrap current profiles by multiplying by a constant \cite{Parisi_2024c}.

Plasma squareness has received less dedicated study compared to other shape parameters. Nonetheless, several benefits of increased $\zeta_0$ have been identified. Ion-scale gyrokinetic simulations of ion-temperature-gradient turbulence suggest that higher $\zeta_0$ improves heat confinement \cite{Joiner2010}, while experiments on DIII-D and MAST-U demonstrate that higher $\zeta_0$ typically allows increased edge plasma pressure \cite{Ferron2000,Lao2001,Leonard2007,Nelson2022,Imada2024} and improved core confinement \cite{Holcomb2009}. ST design studies have shown that moderate positive $\zeta_0$ stabilizes kink modes, giving a 10\% increase in core plasma pressure \cite{Mau1999,Turnbull1999,Jardin2003}. Although this benefit was discounted due to the required increase in poloidal field coil current \cite{Jardin2003,Bromberg2003}, the results of this study suggest that lower $\zeta_0$ values could alleviate these challenges, making the tradeoff more favorable.

Control of $\zeta_0$ is well established \cite{Turnbull1999,Gates2006,Kolemen2011,Ariola2016,Degrave2022} and is often less disruptive than other shape parameters, as plasma x-points and maximum width and height can be maintained. However, adjustments to $\zeta_0$ or other parameters, such as elongation $\kappa_0$, may necessitate changes to internal inductance and related plasma parameters \cite{Menard2016}. Recent work has shown that at low aspect ratio, increased squareness can significantly increase the total fusion power by increasing the burn volume \cite{Parisi_2025a}.

In \Cref{fig:squarenesseq_effxMASTU}, we show several more effects of changing the plasma squareness for MAST-U 48339. An important effect is that the flux expansion is stronger at the outboard midplane (OMP) for higher squareness -- this effect is shown in \Cref{fig:squarenesseq_effxMASTU}(a) where the flux surfaces $\psi_N = [0.3, 0.6, 0.9, 1.0]$ are plotted for different squareness values. The strong flux expansion for higher squareness can be seen by the flux surfaces being more tightly packed for higher squareness. A related result is that the poloidal magnetic field $\mathbf{B}_P =  \nabla \varphi \times \nabla \psi$ increases with higher flux expansion. Here, $\varphi$ is the toroidal angle and $\psi$ is the poloidal flux divided by $2\pi$. Therefore, $B_P \sim |\nabla \psi|/R$ where $R$ is the major radius. Thus, we expect a stronger poloidal field at higher squareness -- this is shown in \Cref{fig:squarenesseq_effxMASTU}(b) where $B_P$ evaluated at the OMP -- $B_{P,\mathrm{out}}$ -- is plotted versus the normalized poloidal flux $\psi_N$, and is larger near the plasma edge for higher squareness. The flux expansion at the OMP, $|\nabla \psi|^2_{\mathrm{out}}$, is plotted in \Cref{fig:squarenesseq_effxMASTU}(c).

A result of the strong outboard flux expansion at higher squareness is that in the edge pedestal region, the field lines have a steeper pitch angle at the OMP, $\chi \equiv \arctan \left( B_{P,\mathrm{out} } / B_{T,\mathrm{out} }  \right) $ where $B_{T,\mathrm{out} }$ is the toroidal field at the OMP. We plot $\chi$ in \Cref{fig:squarenesseq_effxMASTU}(d) for the four squareness values. We also plot the plasma squareness $\zeta$ and elongation $\kappa$ versus $\psi_N$ in \Cref{fig:squarenesseq_effxMASTU}(e),(f). The magnitude of $\zeta$ decreasing with smaller $\psi_N$ is consistent with previous studies showing how shaping tends to decrease at smaller $\psi_N$ values \cite{Atanasiu2024,Freidberg2014,Ball2015}. In agreement with \cite{Parisi_2025a}, plasmas with higher squareness have higher core elongation, boosting the total fusion power.

\begin{table}
\caption{Squareness scans performed in MAST-U and NSTX.}
\begin{ruledtabular}
\centering
  \begin{tabular}{ccc}
   Device & $\zeta_0$ Values & $I_p /$kA \\
    \hline
    MAST-U & -0.28,-0.12,0.04,0.17 & 755 \\
    NSTX & -0.03,0.01,0.13,0.21,0.29,0.39 & 959 \\
  \end{tabular}
\end{ruledtabular}
\label{tab:tabMASTUNSTX}
\end{table}

\section{Gyrokinetics} \label{sec:GK_intro}

In this section, we introduce the gyrokinetic formalism and details of the gyrokinetic simulations that we perform in the pedestal region. There has been substantial work on gyrokinetic simulations and modeling in the pedestal \cite{Told2008,Wang2012,Dickinson2012,Saarelma2013,Canik2013,Fulton2014,Hatch2016,Parisi2020,Hatch2021,Belli2022,Chapman2022,Leppin2023,Walker2023,Predebon2023,Chen2023,Li_2024,Hatch2024,Farcas2024,Jian2024,Dominski2024}. We will focus on the kinetic-ballooning-mode (KBM) \cite{Tang1980,Hastie1981,Dong1999,Pueschel2008,Coury2016,Ma2017,Aleynikova2018}, which is the kinetic analogue to the ideal ballooning mode \cite{Connor1979}. The KBM is thought to be a hard limit on the pedestal pressure gradient in the pedestal center \cite{Snyder2009,Dickinson2012}, and its stability is sensitive to parameters such as the Shafranov shift and magnetic shear. We will use gyrokinetic simulations to predict how plasma shaping affects KBM stability and the pedestal width-height scaling.
KBM stability can be calculated by solving the electromagnetic linear gyrokinetic equation \cite{Catto1978,Frieman1982,Sugama_1998,Parra2008,Abel2013},
\begin{equation}
    \begin{aligned}
    & \frac{ \omega h_{s} + i  v_{\parallel} (\hat{\mathbf{ b} } \cdot \nabla \theta) \partial h_s / \partial \theta - \mathbf{v}_{Ms} \cdot \mathbf{k}_{\perp} h_s - i C_s^l } { \omega - \omega_{*s} } \\
    & = \left( \left( \phi^{tb} - \frac{v_{\parallel}  A_{\parallel}^{tb}}{c} \right) J_{0} + \frac{J_{1}}{b_s} \frac{v_{\perp}^2 B_{\parallel}^{tb}}{c \Omega_s} \right) \frac{Z_s e F_{Ms}}{T_s},
    \end{aligned}
    \label{eq:gke}
\end{equation}
coupled with Maxwell's equations. We solve this system of equations numerically using the gyrokinetic codes GS2 and CGYRO \cite{Dorland2000,Candy2016} -- in this work we present results from GS2. Here, $\omega$ is the complex mode frequency, $h_s$ is the non-adiabatic distribution function for species $s$, $v_{\parallel}$ and $v_{\perp}$ are the parallel and perpendicular components of the particle velocity, $\hat{\mathbf{ b} } =  \mathbf{B} /B$ is the magnetic field unit vector where $\mathbf{B}$ is the equilibrium magnetic field, $\theta$ is a poloidal angle, $\mathbf{v}_{Ms}$ is the magnetic drift for species $s$, $\mathbf{k}_{\perp} = k_\psi \nabla \psi + k_{\alpha} \nabla \alpha$ is the perpendicular wavenumber where $\psi$ and $\alpha$ are flux-tube spatial coordinates described in \Cref{sec:geodefinitions}, $C_s^l$ is a linearized collision operator, $\phi ^{tb}$, $A_{\parallel}^{tb}$, and $B_{\parallel}^{tb}$ are the fluctuating electrostatic potential, parallel vector potential, and parallel magnetic field, $c$ is the speed of light, $J_0$ and $J_1$ are Bessel functions of the first kind, $b_s = k_{\perp} v_{\perp} / \Omega_s$ where $\Omega_s$ is the cyclotron frequency for species $s$, $Z_s$ is the species charge normalized to the proton charge $e$, $F_{Ms}$ is the equilibrium Maxwellian distribution function, and $T_s$ is the temperature for species $s$. The perpendicular radial E $\times$ B drift frequency for species $s$ is
\begin{equation}
\omega_{*s} = \frac{c}{B} \frac{T_{s}}{Z_s eL_{ns}} k_y \left( 1 + \eta_s \left( \frac{m_s v^2}{2 T_s} - \frac{3}{2} \right)  \right),
\end{equation}
where $L_{ns} = - \left( \partial/\partial r \ln n_s \right)^{-1}$, the binormal wavenumber is $k_y \equiv - \left( \hat{\mathbf{ b} } \times \nabla r \right) \cdot \mathbf{k}_{\perp}$, the ratio of the density to temperature gradient is $\eta_s \equiv L_{ns}/L_{Ts}$, $m_s$ is the particle mass, and $v$ is the particle speed.

We will calculate the KBM growth rate across pedestal width and height for NSTX 132543. As in previous works \cite{Parisi_2024,Parisi_2024b,Parisi_2024c}, the pedestal width and height is varied self-consistently with equilibrium reconstruction starting from the experimental value at 614ms. In this work, we choose to vary the pedestal pressure at fixed density \cite{Parisi_2024}. Linear initial value electromagnetic collisional flux-tube gyrokinetic simulations are performed using GS2 \cite{Dorland2000} at seven flux surfaces spanning the pedestal half-width with binormal wavenumbers $k_y \rho_i \in [0.06, 0.12, 0.18]$ and ballooning wavenumber angle $\theta_0 = 0.0$. The magnetic geometry is calculated numerically from the plasma equilibrium. For each simulation, the fastest growing mode is identified with an automated mode finder algorithm \cite{Parisi_2024} based on the mode `fingerprints' \cite{Kotschenreuther2019}. For the six NSTX equilibria with different squareness values, a pedestal width and height scan, the radial grid with seven points and three binormal wavenumbers, this corresponds to a database of 15,246 gyrokinetic simulations.

For more definitions of pedestal width and height as well as details on how the equilibrium variation is performed, see \Cref{sec:peddefinitions} and \cite{Parisi_2024,Parisi_2024b,Parisi_2024c}. We now proceed to calculate the pedestal width-height scaling across squareness.

\section{Gyrokinetic Width-Height Scaling} \label{sec:GKwidthheight}

\begin{figure}[!tb]
    \centering
    \begin{subfigure}[t]{0.45\textwidth}
    \centering
    \includegraphics[width=1.0\textwidth]{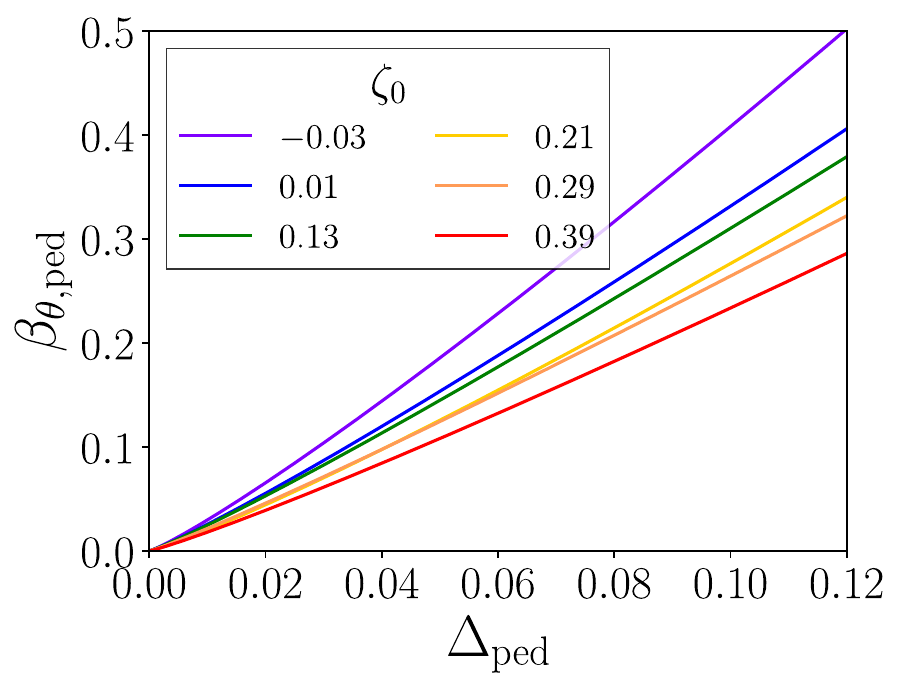}
    \caption{}
    \end{subfigure}
    \begin{subfigure}[t]{0.45\textwidth}
    \centering
    \includegraphics[width=1.0\textwidth]{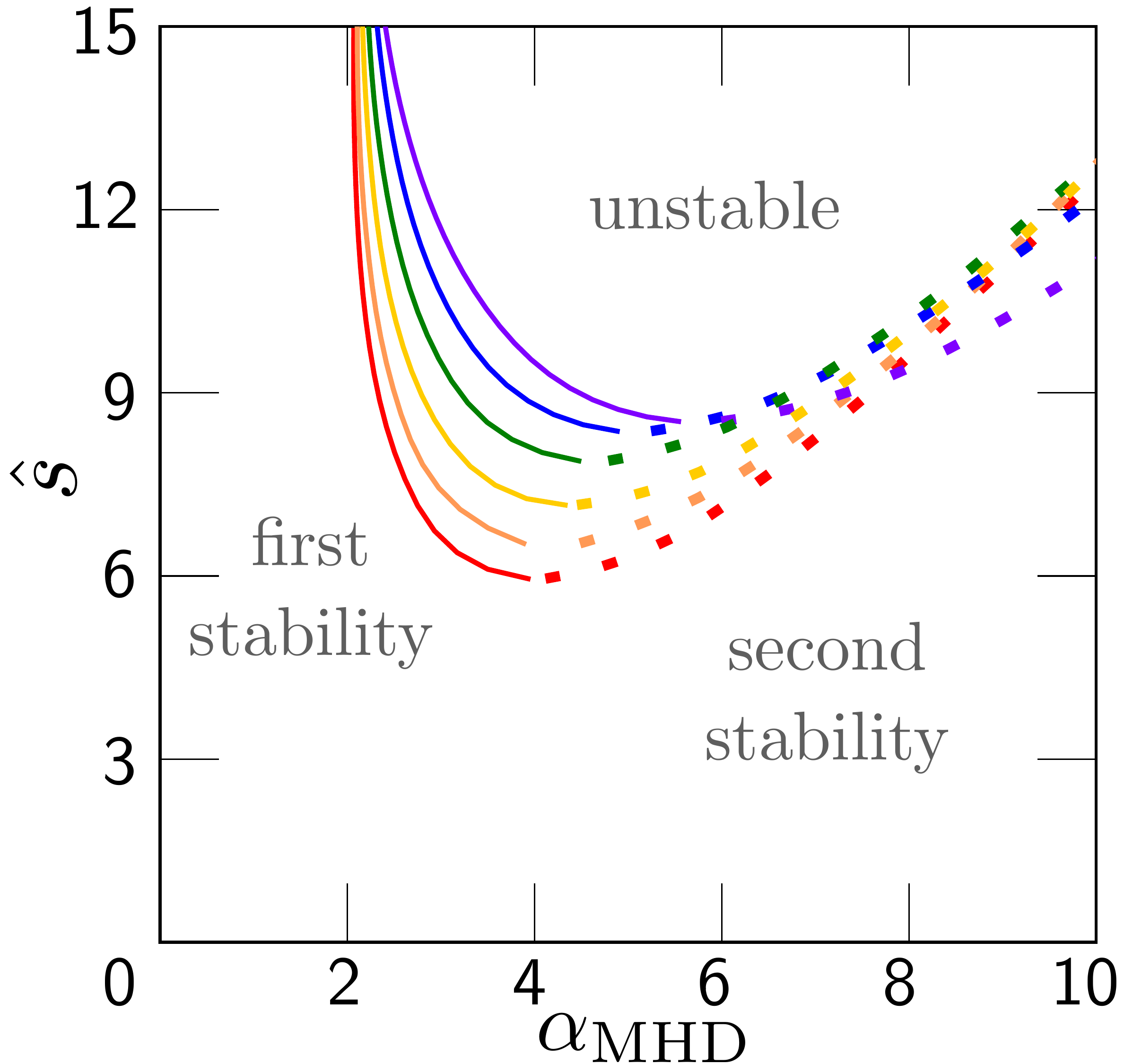}
    \caption{}
    \end{subfigure}
     ~
    \caption{(a): Gyrokinetic pedestal width-height scalings for NSTX 132543 with six squareness values. (b): $\hat{s}-\alpha_\mathrm{MHD}$ scan at the mid-pedestal radius for the six squareness values.}
    \label{fig:squareness-scan}
\end{figure}

In this section, we show the effect of plasma squareness on the width-height scaling. 

The pedestal width-height scaling is a constraint for the pedestal width and height calculated from ballooning stability \cite{Snyder2009}. In this work, we calculate the width-height scaling from KBM stability obtained from gyrokinetic simulations. The width-height scaling has two branches -- wide and narrow \cite{Parisi_2024b} -- corresponding closely to first and second ballooning stability respectively. Because ST H-mode pedestals are often in first ballooning stability \cite{Parisi_2024c,Nelson2024c}, and because of the potential ELM-free benefits of being in the first stability region, in this work we focus on the wide pedestal branch. In the wide branch, the width-height scaling gives the highest possible pedestal pressure for a given pedestal width for KBM-limited profiles. The pedestal pressure is often expressed by the normalized quantity $\beta_{\theta, \mathrm{ped}}$ and the width $\Delta_{\mathrm{ped} }$; see \Cref{sec:peddefinitions} for exact definitions. Therefore, the width-height scaling is found by calculating the coefficients $a$ and $b$ in $\Delta_{\mathrm{ped} } = a \left( \beta_{\theta, \mathrm{ped}} \right)^b$.

We find the width-height scalings for the first stability branch for NSTX equilibria with six squareness values, shown in \Cref{fig:squareness-scan}(a). Higher plasma squareness gives a lower maximum characteristic gradient $\sim \beta_{\theta, \mathrm{ped}} / \Delta_{\mathrm{ped} }$. We will show that equilibria with pedestals with lower average gradients could be useful for ELM avoidance because they move the pedestal further from the ELM boundary.

As a consistency check for the gyrokinetic width-height scalings in \Cref{fig:squareness-scan}(a), we also calculate the infinite toroidal mode number (infinte-$n$) ballooning stability \cite{Connor1979} boundaries across the NSTX equilibria with six different squareness values. We use the BALOO code \cite{Miller1997}. In \Cref{fig:squareness-scan}(b), we plot the $\hat{s}-\alpha_\mathrm{MHD}$ stability boundaries for the infinite-$n$ ballooning mode \cite{Greene1981}. Here, the Shafranov shift
\begin{equation}
\alpha_\mathrm{MHD} = (2/\pi)(\partial V / \partial \psi) \sqrt{V/(2\pi^2 R_0)} (dp/d\psi),
\end{equation}
is destabilizing at lower values where pressure gradients overcome the stabilizing effects of field-line bending \cite{Connor1979}, $R_0$ is the plasma major radius, $V$ is the volume enclosed in the flux surface, and $p$ is the total plasma pressure. At higher $\alpha_\mathrm{MHD}$ values, the local magnetic shear at the outboard midplane becomes large in magnitude, leading to second stability \cite{Greene1981}. The global magnetic shear
\begin{equation}
\hat{s} = (\psi / q) dq / d\psi,
\end{equation}
measures the radial derivative of the safety factor $q$.

The infinite-$n$ stability calculations are performed on the same flux-surface value for equilibria with identical $\beta_{\theta, \mathrm{ped}}$ and $\Delta_{\mathrm{ped} }$. Increasing the squareness decreases the critical magnetic shear and $\alpha_\mathrm{MHD}$ for instability. Our infinite-$n$ ballooning calculation is consistent with the gyrokinetic width-height scalings in \Cref{fig:squareness-scan}(a), showing that increasing plasma squareness is destabilizing for the ballooning mode in first stability.

The width-height scalings in \Cref{fig:squareness-scan}(a) predict that different squareness values could change the maximum achievable $\beta_{\theta, \mathrm{ped}}$ by a factor of $\sim$1.5. In the next section, we study the combined effects of squareness on both KBM and PBM stability for MAST-U.

\section{Combined Kinetic and Peeling-Ballooning Constraints} \label{sec:GK_peeling_widthheight}

In this section, we study the combined constraints for kinetic-ballooning and peeling-ballooning modes for MAST-U discharge 48339 at 600ms with varied squareness. We first present the peeling-ballooning stability calculations, and then combine them with kinetic-ballooning stability.

\subsection{Peeling-Ballooning Stability}

We study ideal PBM stability across plasma squareness using the ELITE code \cite{Wilson2004}. We run ELITE on each equilibrium in $\Delta_{\mathrm{ped}}$, $\beta_{\theta,\mathrm{ped}}$ space for toroidal mode numbers $n \in [5,10,15,20,25]$ and a range of poloidal mode numbers. We then find the fastest growing mode over all mode numbers in units of $\gamma / (\omega_{*i} /4)$, where $\gamma$ is the linear PBM growth rate and $\omega_{*i}$ is the diamagnetic frequency. The PBM is considered unstable when $\gamma \gtrsim \omega_{*i} /4$ \cite{Snyder2009b}.

\begin{figure}[!tb]
    \centering
    \begin{subfigure}[t]{0.49\textwidth}
    \centering
    \includegraphics[width=1.0\textwidth]{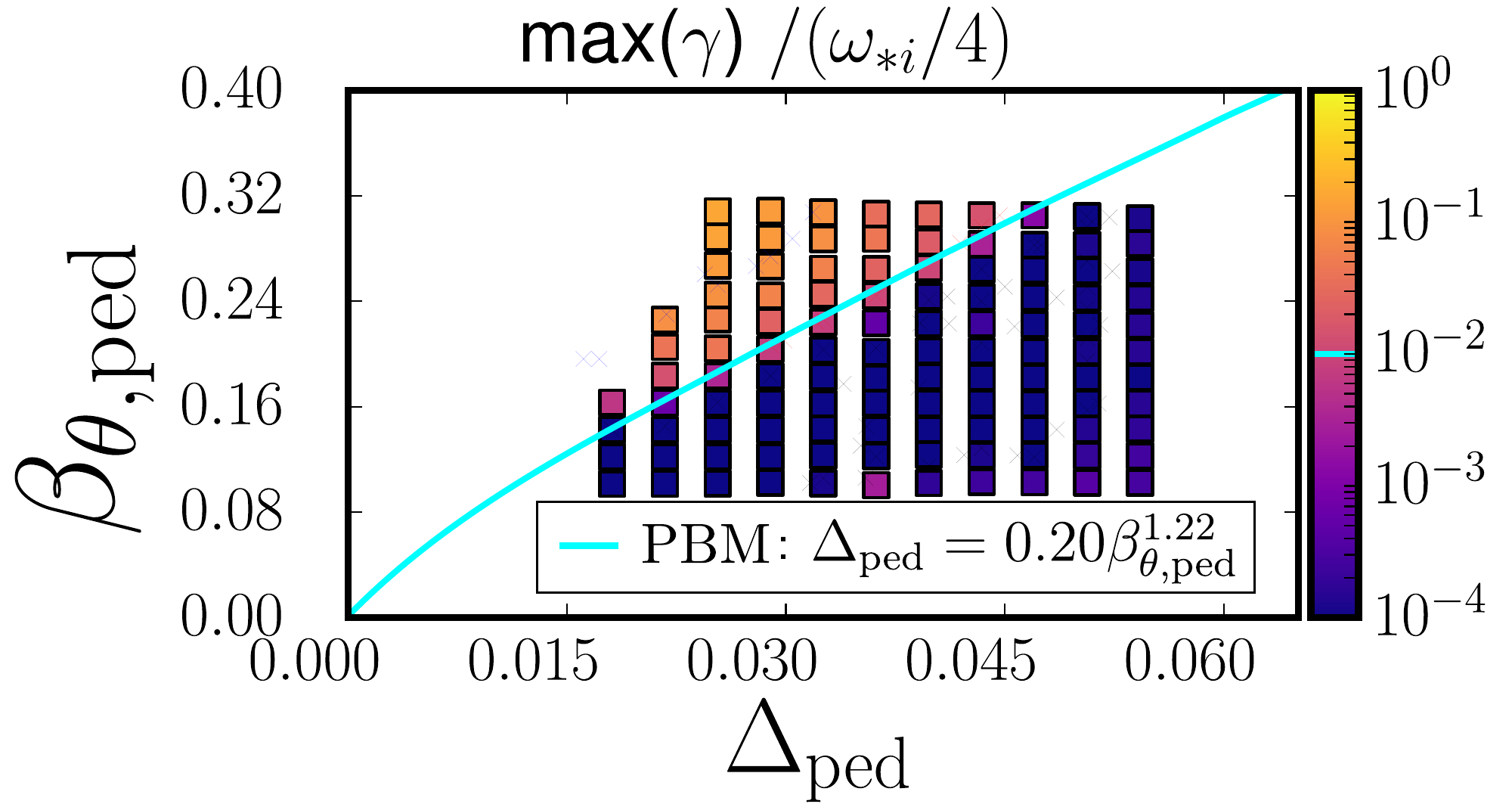}
    \caption{}
    \end{subfigure}
    \centering
    \begin{subfigure}[t]{0.49\textwidth}
    \centering
    \includegraphics[width=1.0\textwidth]{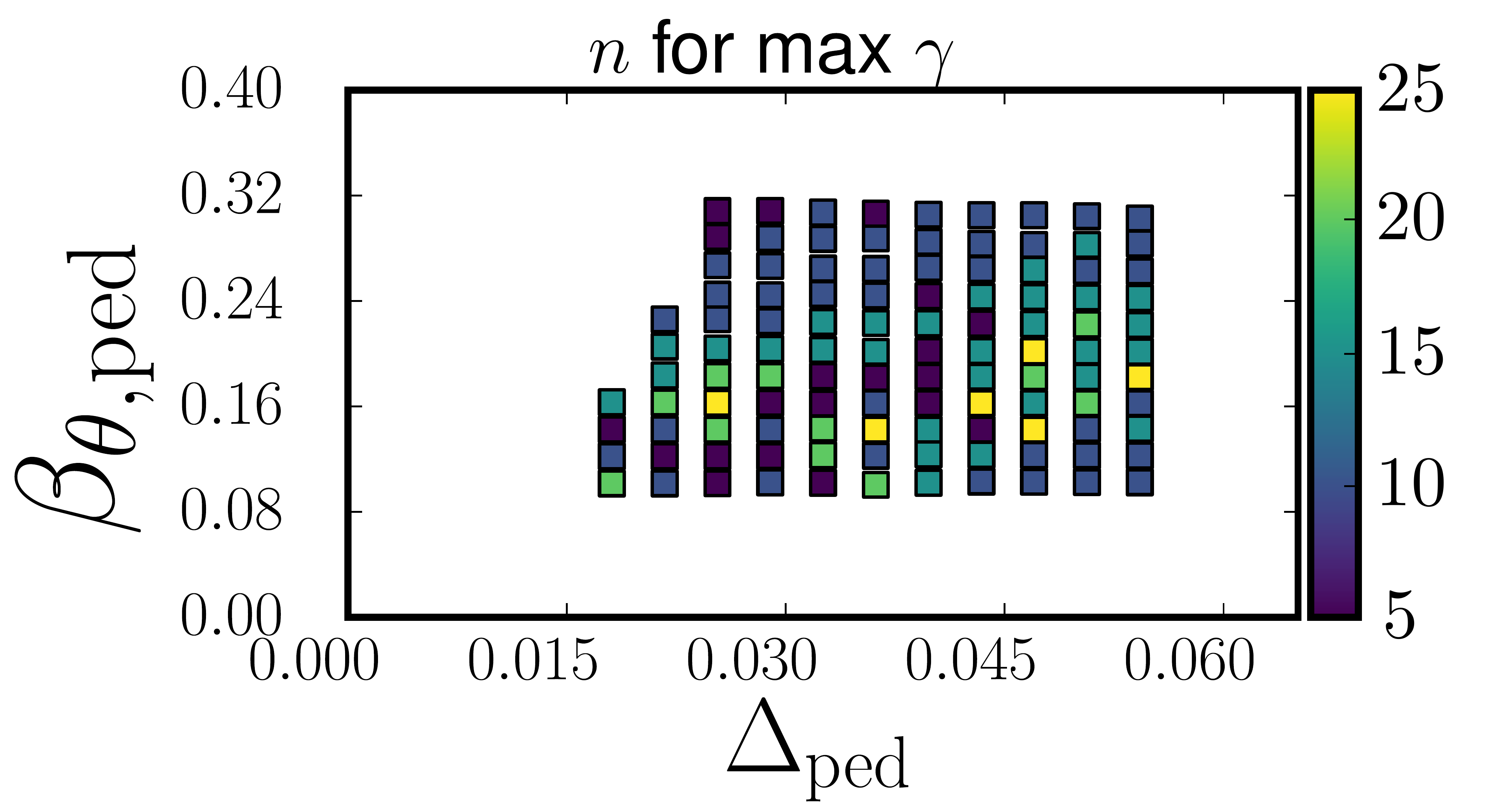}
    \caption{}
    \end{subfigure}
    \caption{Ideal PBM stability across pedestal width and height for MAST-U 48339 with $\zeta_0 = 0.17$. (a) Fastest growing mode. (b) Mode number $n$ for fastest growing mode.}
    \label{fig:ELITE_growth}
\end{figure}

It is well-known that capturing ideal-PBM stability at lower aspect ratio is challenging \cite{Kleiner2021,Imada2024}. In \Cref{fig:ELITE_growth}(a), we plot the fastest growing modes across pedestal width and height for MAST-U 48399 with $\zeta_0 = 0.17$. Notably, there is no region where $\gamma > \omega_{*i} /4$. This contradicts the experimental discharge, which is close to the Type-1 ELM boundary. Therefore, to approximately capture the ELM-limit, we use a `fine-tuned' criterion for the ideal PBM stability boundary as performed in \cite{Imada2024,Imada2024b} for MAST-U, which is $\gamma > \gamma_f$, where we choose $\gamma_f$ based on approximately where the ELM-limit is for the experiment. In this work, we choose $\gamma_f = \omega_{*i} /400$ to approximate the experimental point. We fit the fine-tuned width stability boundary ($\gamma_f = \omega_{*i} /400$) in \Cref{fig:ELITE_growth}(a) to a power law $\Delta_{\mathrm{ped} } = a \left( \beta_{\theta, \mathrm{ped} } \right)^b$, where $a$ and $b$ are fitting parameters. We denote this curve as the `PBM width-height scaling' \cite{Snyder2004,Snyder2009b}. We find the PBM scaling
\begin{equation}
\Delta_{\mathrm{ped} } = 0.20 \left( \beta_{\theta, \mathrm{ped} } \right)^{1.22},
\end{equation}
which is in good agreement with the theoretical scaling of $\Delta_{\mathrm{ped} } \sim \left( p_{\mathrm{ped} } \right)^{4/3}$ \cite{Snyder2004,Snyder2009b,Snyder2011}. Note that this technique of fine-tuning the critical $\gamma_f$ is analogous to the technique in \cite{Imada2024,Imada2024b}, but differs in that we normalize the growth rate by $\omega_{*i}$, whereas \cite{Imada2024,Imada2024b} normalized the growth rate by the Alfven frequency. We plot the toroidal mode number $n$ corresponding to the fastest growing mode in \Cref{fig:ELITE_growth}(b) -- as the pressure gradient increases, the mode number generally decreases. Around the PBM stability boundary, the fast-growing mode number is typically $n = 10 - 15$.

When performing the same exercise to calculate the PBM scaling expression on NSTX discharge 132543 across pedestal width and height, we found that $\gamma$ is vanishingly small everywhere in pedestal width-height space, which is consistent with ideal PBM calculations in \cite{Kleiner2021}. Due to the difficulty of calculating ideal PBM stability in NSTX, we only focus on ideal PBM stability in MAST-U. In the next subsection, we calculate the PBM scaling across different squareness values for MAST-U 48339.

\subsection{Combined Kinetic and Peeling-Ballooning Constraints}

\begin{figure}[bt]
    \centering
    \includegraphics[width=0.5\textwidth]{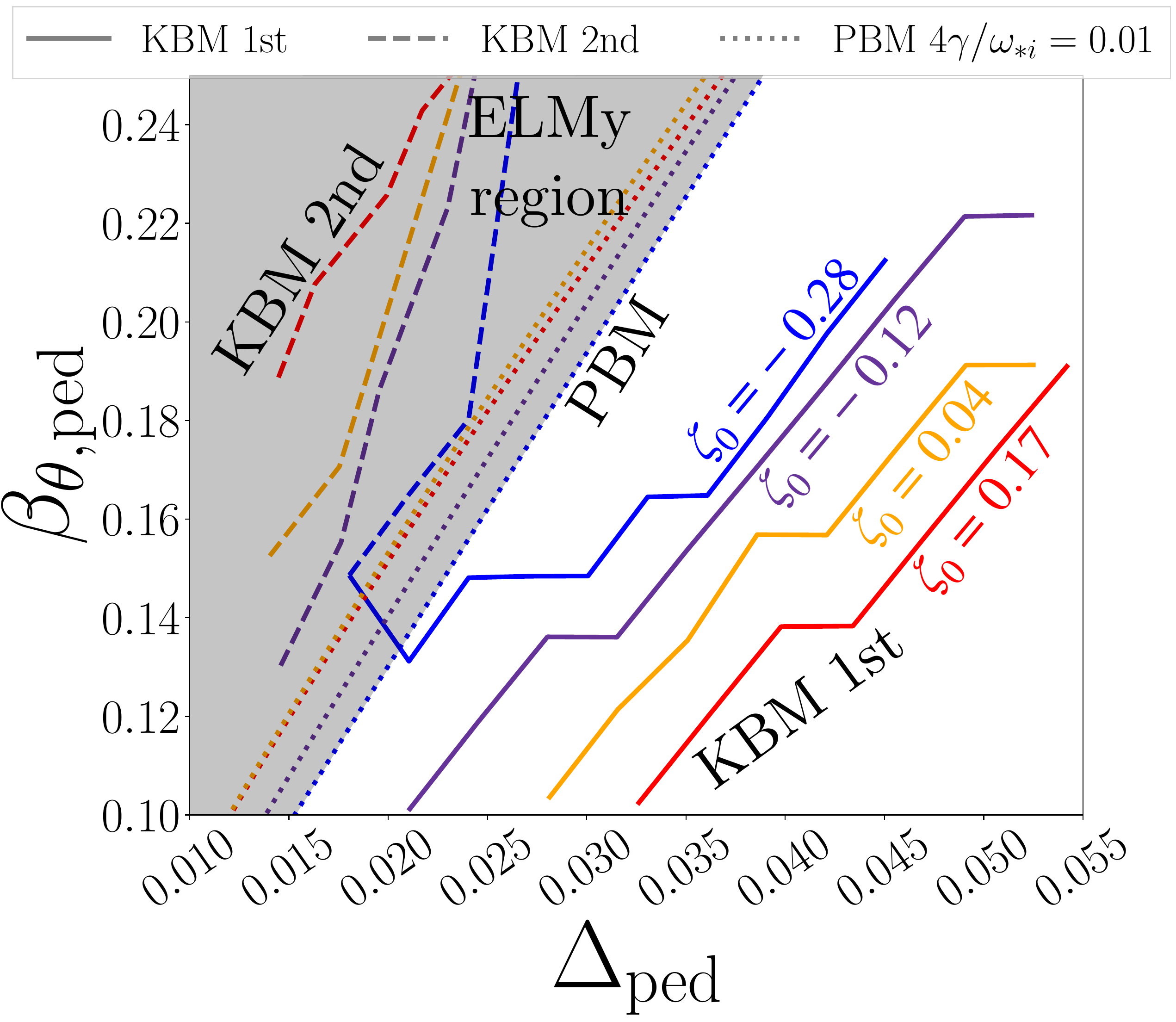}
    \caption{First and second KBM stability boundaries and PBM stability boundaries across four squareness values in MAST-U 48339.}
    \label{fig:squareness_first_second_stab_MASTU}
\end{figure}

We now combine the KBM constraint -- obtained from gyrokinetic simulations -- and the PBM constraint -- obtained from ELITE simulations in the previous section -- for MAST-U 48339 across plasma squareness.

We first describe the effects of plasma squareness on the stability boundaries for KBM for MAST-U 48339 (whereas the previous section was for NSTX 132543). Shown in \Cref{fig:squareness_first_second_stab_MASTU}, the solid lines show the KBM first stability boundary for MAST-U 48399. Similar to our results for NSTX, for MAST-U 43889, in first stability, the KBM is stabilized with decreasing squareness. Increasing squareness closes off second stability access, shown by the dashed lines in \Cref{fig:squareness_first_second_stab_MASTU}. If the pedestal remained KBM first stability limited, an increase in squareness from -0.28 to 0.17 would cause significant pedestal widening at fixed $\beta_{\theta, \mathrm{ped} }$ because the maximum critical KBM gradient decreases with higher squareness. The closing off of second stability is also partially consistent with findings (at higher aspect ratio) in \cite{Nelson2022}. However, because we are interested in ELM-free discharges, we will not study second stability further.

A PBM scaling expression is calculated for the four equilibria with different squareness values using ELITE in the same manner presented in \Cref{fig:ELITE_growth}. While the KBM stability boundary is sensitive to plasma squareness, the PBM stability boundary is insensitive to squareness, as shown by the dotted lines in \Cref{fig:squareness_first_second_stab_MASTU}. 

As a numerical experiment, we found that we could only change the PBM boundary with unphysically negative or large squareness (where the plasma last closed flux surface is touching the vessel). The physical mechanism for the relative invariance of PBM stability across plasma squareness in MAST-U is still an outstanding question, but may be related to a `decoupling' of peeling-ballooning and high-n ballooning modes \cite{Imada2024b}. Another explanation could be that extended-MHD effects are missing \cite{Nystrom2022,Kleiner2022} -- this could lead to different growth rates at different squareness. Physically, this could result from an $m = 8$ shaping moment (outer squareness) not affecting PBMs as much as KBMs due to the lower mode number of the PBMs. Note that our result on the relative invariance of PBM stability to plasma shaping may seem to contrast with \cite{Merle2017}, where a negative triangularity TCV ELMy H-mode had an ELM stability boundary at least twice lower pedestal height of its positive triangularity counterpart. However, our results may not contradict this because we are studying a much higher shaping moment ($m = 8$ versus $m = 3$ in \cite{Merle2017}), a lower aspect ratio, and many other discharge parameters vary.

In summary, we have calculated combined KBM and ideal PBM pedestal width-height scalings for four MAST-U equilibria with different plasma squareness. Shown in \Cref{fig:squareness_first_second_stab_MASTU}, the ideal PBM scaling barely changes with squareness, but the KBM scaling is sensitive to the plasma squareness.

\section{Random Forest Model} \label{sec:random_tree_model}

In this section, we will introduce a Random Forest (RF) model for (a) the KBM growth rates for NSTX and (b) the pedestal height distance from the KBM first stability boundary for MAST-U. The inputs to the RF model will be geometric coefficients appearing in the gyrokinetic equation.

Machine learning is widely used in fusion research, being applied to equilibrium reconstruction \cite{Felici2011,Lao2022,Candido2023}, disruption prediction and mitigation,  \cite{Rea2019,Ho2019,Vega2022,Sabbagh2023,Gambrioli_2025}, profile prediction \cite{Boyer2021,Abbate2021,Dubbioso2023}, turbulence simulation acceleration \cite{Ma2020,Van2020}, plasma control \cite{Degrave2022}, plasma heating modeling \cite{Wallace2022,Sanchez2024}, stability assessment \cite{Piccione2020,Piccione2022} and more \cite{Smith2013,Meneghini2021,Kit2023,vanLeeuwen_2025,Pavone2023,Landreman2025,Parisi2025c}.

RF models are a class of machine learning algorithms, widely recognized for their predictive accuracy, resistance to overfitting, and interpretability \cite{Breiman2001,Liaw2002,Cutler2007}. RFs are a method based on decision trees, which are constructed to partition data into regions of similar output values by iteratively splitting the dataset along input feature thresholds. In this work, the input features are magnetic geometry coefficients. While a single decision tree is prone to overfitting—overly conforming to training data and performing poorly on unseen data—the RF algorithm mitigates this issue by averaging predictions across many trees.

The construction of an RF begins by training multiple decision trees on samples, where each sample is drawn with replacement from the original dataset. This ensures that individual trees are exposed to slightly different data. Further randomness is introduced at each decision node: instead of evaluating all features to determine the best split, RF restricts the search to a randomly selected subset of features. This additional randomness reduces correlations among trees.

Once trained, the RF makes predictions by aggregating the outputs of all trees in the ensemble. For classification tasks, the output is determined by majority voting among the trees, while for regression, the final prediction is the average of the individual tree outputs \cite{Breiman2001,Pal2005}.

RF models are particularly useful for capturing complex nonlinear relationships and interactions among input features, making them well-suited for the high-dimensional datasets in this work \cite{Breiman2001,Geurts2006}. Moreover, RFs provide interpretable metrics such as feature importance and permutation importance, which quantify the relative contribution of each input feature to the model’s predictions. While we also analyzed the data in this work using other machine learning algorithms, the RF algorithm consistently performed the best.

\subsection{Gyrokinetics and Geometry}

\begin{figure*}
    \centering
    \begin{subfigure}[t]{0.69\textwidth}
    \centering
    \includegraphics[width=1.0\textwidth]{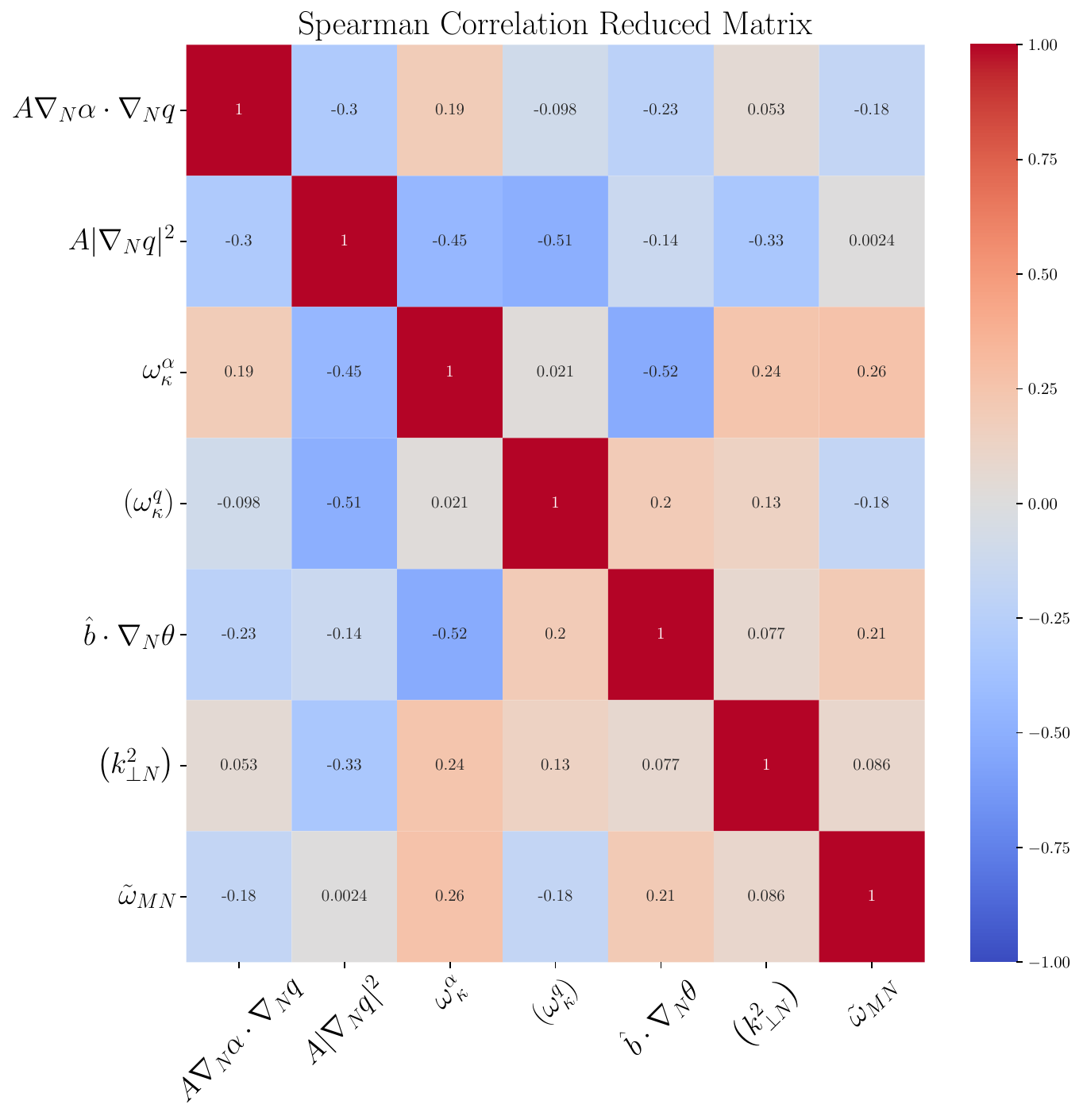}
    \end{subfigure}
    \caption{Spearman correlation of geometric coefficients in NSTX pedestals with six squareness values based on NSTX 132543. Coefficients that have an absolute Spearman value greater than 0.70 were removed from model analysis -- the full matrix before strongly correlated features are removed is shown in \Cref{fig:spearman_NSTX_full}.}
    \label{fig:spearman_NSTX}
\end{figure*}

\begin{table*}[tb]
\caption{Geometry quantities.}
\begin{ruledtabular}
\centering
  \begin{tabular}{ccp{9cm}}  %
   Name & Quantity & Description \\
    \hline
    Parallel streaming & $\hat{\mathbf{ b} } \cdot \nabla_N \theta$ & 
    Measures particle motion along field lines through poloidal angle $\theta$; larger values typically entail stronger Landau damping. \\
    
    Inter-field distance & $A |\nabla_N \alpha|^2$ & 
    Describes spacing between field lines on a flux surface; larger values correspond to tighter packing and stronger binormal-to-perpendicular wavenumber mapping. \\
    
    Field line twist & $A \nabla_N \alpha \cdot \nabla_N q$ & 
    Quantifies how the field line label changes with radius via $q$; controls shearing between field lines. \\
    
    Flux expansion & $A |\nabla_N q|^2$ & 
    Measures how flux surface spacing grows with radius; high near the X-point and divertor. \\
    
    Perpendicular wavenumber & $k_{\perp N}^2$ & 
    Sets the perpendicular scale of turbulence; higher values correspond to smaller eddies and stronger FLR suppression. \\
    
    Magnetic field strength & $B_N$ & 
    Sets gyrofrequency, Larmor radius, and drift scales; variations in $B_N$ drive grad-B drifts. \\
    
    In-surface grad-B drift frequency & $\omega_{\nabla B}^\alpha$ & 
    Drift from magnetic field gradients projected within the flux surface. \\
    
    In-surface curvature drift frequency & $\omega_{\kappa}^\alpha$ & 
    Drift from field line curvature projected within the flux surface. \\
    
    Radial curvature drift frequency & $\omega_{\kappa}^q$ & 
    Similar to $\omega_{\kappa}^\alpha$ but projected radially. \\
    
    Total magnetic drift frequency & ${\omega}_{M N}$ & 
    Combines grad-B and curvature drifts. \\
    
    Shafranov shift & $\alpha_{\mathrm{MHD} }$ & 
    Outward displacement of flux surfaces due to pressure; modifies magnetic shear and ballooning stability. \\
    
    Local magnetic shear & $|s_{\mathrm{local}}|$ & 
    Measures how quickly field line pitch changes with radius at a given poloidal location; high shear can suppress turbulence. \\
  \end{tabular}
\end{ruledtabular}
\label{tab:tabgeo_quants}
\end{table*}

Before showing the results of the RF model, we describe the preparation of feature inputs, in this case, the magnetic geometry coefficients. In order to meaningfully compare gyrokinetic simulations across different flux surfaces and equilibria, we will study normalized, dimensionless quantities, typically denoted with a `N' subscript. More details on normalization conventions can be found in \Cref{sec:geodefinitions}. The dimensionless, normalized gyrokinetic equation is
\begin{equation}
    \begin{aligned}
    & \frac{ \left[ \omega_N + i v_{t N s} v_{\parallel N} (\hat{\mathbf{ b} } \cdot \nabla_N \theta) \frac{ \partial}{ \partial \theta} - \mathbf{v}_{MNs} \cdot \mathbf{k}_{\perp N}  \right] h_{Ns} - i C_{Ns}^l } { \omega_N - \omega_{*Ns} } \\
    & = \left( \left( \phi^{tb}_N - v_{t N s} v_{\parallel N}  A_{\parallel N}^{tb} \right) J_{0} + \frac{J_{1}}{b_{Ns}} v_{\perp N}^2 B_{\parallel N}^{tb} \right) \frac{Z_s}{T_{Ns}}.
    \end{aligned}
    \label{eq:gke_normalized}
\end{equation}
We normalize quantities to the electron density and temperature and the deuterium mass and charge. This choice means that the electron equation and its geometric coefficients will have the same normalization when comparing different simulations. We will study the gyrokinetic operators appearing in the electron equation for \Cref{eq:gke_normalized},
\begin{equation}
    \begin{aligned}
    & \frac{ \left[ \omega_N + i v_{t N e} v_{\parallel N} (\hat{\mathbf{ b} } \cdot \nabla_N \theta) \frac{ \partial}{ \partial \theta} - \mathbf{v}_{MNe} \cdot \mathbf{k}_{\perp N}  \right] h_{Ne} - i C_{Ne}^l } { \omega_N - \omega_{*Ne} } \\
    & = - \left( \left( \phi^{tb}_N - v_{t N e} v_{\parallel N}  A_{\parallel N}^{tb} \right) J_{0} + \frac{J_{1}}{b_{Ne}} v_{\perp N}^2 B_{\parallel N}^{tb} \right).
    \end{aligned}
    \label{eq:gke_normalized_electron}
\end{equation}
In \Cref{eq:gke_normalized_electron}, there are five main geometric terms: $\hat{\mathbf{ b} } \cdot \nabla_N \theta$, $\omega_{*Ne}$, $k^2_{\perp N}$, $\omega_{M N}$, and $B_N$.

The parallel streaming operator corresponding to particles streaming along field lines is
\begin{equation}
\hat{\mathbf{ b} } \cdot \nabla_N \theta,
\label{eq:parallel_stream}
\end{equation}
the electron E $\times$ B drift frequency is $\omega_{*Ne}$, the perpendicular wavenumber is
\begin{equation}
\begin{aligned}
& k^2_{\perp N} = A k^2_{yN} \left| \left| \nabla_N \alpha \right|^2 + 2 \theta_0 \nabla_N \alpha \cdot \nabla_N q + \theta_0^2 \left| \nabla_N q \right|^2 \right|,
\end{aligned}
\end{equation}
where
\begin{equation}
A \equiv \left( \frac{d\psi_N}{d (r/a)} \right)^2,
\end{equation}
the magnetic drift operator is
\begin{equation}
\mathbf{v}_{MNe} \cdot \mathbf{k}_{\perp N} = - \omega_{M N},
\end{equation}
where
\begin{equation}
\begin{aligned}
& \omega_{M N}  = \\
& \frac{k_{yN}}{2} \left[ \frac{v_{\perp N}^2 }{2} \left( \omega_{\nabla B}^\alpha + \theta_0 \omega_{\kappa}^q \right) +  v_{\parallel N}^2 \left( \omega_{\kappa}^\alpha + \theta_0 \omega_{\kappa}^q \right) \right].
\end{aligned}
\end{equation}
Here, $\theta_0 = k_x / (k_y \hat{s})$ is the ballooning wavenumber angle where $\hat{s} = (r/q)(dq / d r)$ is the magnetic shear, $q$ is the safety factor, $\omega_{\nabla B}^\alpha$ and $\omega_{\kappa}^\alpha$ are the in-surface grad-B and curvature drift frequencies, and $\omega_{\kappa}^q$ is the radial curvature drift frequency (equal to the radial grad-B drift frequency). See \Cref{sec:geodefinitions} for the drift frequency definitions. To simplify the perpendicular and parallel velocity space dependence, we define a new quantity ${\omega}_{M N} $ evaluated with $v_{\perp N}^2 = v_{\parallel N}^2 = 1.0$,
\begin{equation}
{ \omega}_{M N} = \frac{k_{yN}}{4} \left[  \omega_{\nabla B}^\alpha + 3 \theta_0 \omega_{\kappa}^q + 2 \omega_{\kappa}^\alpha \right].
\end{equation}
The final geometric parameter is the total magnetic field strength
\begin{equation}
B_N = B / B_r,
\end{equation}
which varies with poloidal angle on a flux surface. Because of its particular importance for KBM stability, in our stability analysis we will use the Shafranov shift parameter $\alpha_{\mathrm{MHD} }$ instead of $\omega_{*Ne}$. For simplicity, we neglect the role of collisions. We also include the local magnetic shear in our analysis \cite{Greene1981,Gaur2023}
\begin{equation}
s_{\mathrm{local}} = - \mathbf{B} \cdot \nabla \left( \frac{\nabla \alpha \cdot \nabla \psi}{|\nabla \psi|^2} \right).
\label{eq:slocal}
\end{equation}
For infinite-$n$ ideal ballooning modes the local magnetic shear stabilizing effect is proportional to the absolute value of $s_{\mathrm{local}}$ \cite{Freidberg2014}. Because $s_{\mathrm{local}}$ decreases and passes through zero as the pedestal pressure gradients increase, we will study $|s_{\mathrm{local}}|$ rather than $s_{\mathrm{local}}$.

With the exception of $\alpha_{\mathrm{MHD} }$, all of the geometric coefficients have a $\theta$ dependence. While gyrokinetic solvers account for this $\theta$ dependence, it would be challenging to include all of the gridpoints in $\theta$ for the geometric terms in a ML model. This is because it would introduce far too many parameters. Therefore as a simplification, we will exploit the fact that KBM eigenmode usually peaks at the OMP at $\theta = 0.0$ and only evaluate the OMP geometric coefficients. In \Cref{sec:eigen_average}, we show an alternative approach by averaging geometric coefficients over the KBM eigenmode. We show that the simpler approach with coefficients at the OMP is only marginally less accurate than using the eigenmode-averaged coefficients. For a predictive model, evaluating the coefficients at the OMP has the advantage that no eigenmode model is required to predict the growth rate -- only the geometric coefficients at the OMP.

Because we are interested in predicting the KBM growth rate, which depends on the binormal wavenumber $k_y$, we will consider both the geometric coefficients that are independent of $k_y$, such as $|\nabla_N \alpha|^2$, as well as the operators that depend on $k_y$ such as $k^2_{\perp N}$. Therefore, for our analysis of the effect of geometry on the KBM growth rate, we include all of the following quantities, evaluated at the OMP:
\begin{equation}
\begin{aligned}
& \hat{\mathbf{ b} } \cdot \nabla_N \theta, \; A |\nabla_N \alpha|^2,\; A \nabla_N \alpha \cdot \nabla_N q ,\; A |\nabla_N q|^2, k_{\perp N}^2, \\
& B_N, \omega_{\nabla B}^\alpha, \; \omega_{\kappa}^q, \; \omega_{\kappa}^\alpha, \; {\omega}_{M N}, \; \alpha_{\mathrm{MHD} }, \; |s_{\mathrm{local}}|.
\end{aligned}
\label{eq:OMP_features}
\end{equation}
Our task is to determine how these geometric quantities affect gyrokinetic stability, and how plasma shaping, in this case plasma squareness, affects these geometric quantities.

These quantities are listed in \Cref{tab:tabgeo_quants}. The physical intuition behind each of these terms is as follows: $\hat{\mathbf{ b} } \cdot \nabla_N \theta$ describes how particles stream along magnetic field lines as they move through poloidal angle $\theta$. If field lines loop quickly around poloidally, this term is large, meaning fast parallel motion in the poloidal direction — important for Landau damping. $A |\nabla_N \alpha|^2$ measures how tightly packed the magnetic field lines are within a flux surface — specifically, the spacing between neighboring field lines labeled by the field line label $\alpha$. Smaller spacing (larger $A |\nabla_N \alpha|^2$) means a given binormal wavenumber $k_{yN}$ results in a larger perpendicular wavenumber $k_{\perp N}$. The field line twist $A \nabla_N \alpha \cdot \nabla_N q$ captures how the field line label $\alpha$ changes moving radially outward (through changes in the safety factor $q$). More twist means more shearing between neighboring field lines — which can stabilize or destabilize turbulence depending on the flux-surface alignment. The flux expansion $A |\nabla_N q|^2$ describes how stretched out flux surfaces are radially — when the safety factor changes rapidly in space, the spacing between flux surfaces grows. High flux expansion is often seen near the X-point and divertor. The perpendicular wavenumber $k_{\perp N}^2$ gives the square of the perpendicular size of turbulent eddies. Higher $k_{\perp N}^2$ means smaller turbulent structures, more affected by finite Larmor radius (FLR) effects, which can decrease turbulent amplitudes and lower growth rates. The magnetic field strength $B_N$ determines gyrofrequency, Larmor radius, and drift speeds. Stronger $B_N$ means small Larmor radius and reduced perpendicular transport. Spatial variation in $B_N$ is also what drives grad-B drifts. The in-surface grad-B drift frequency $\omega_{\nabla B}^\alpha$ arises from the gradient in magnetic field strength, but projected within the flux surface, in the $\alpha$ direction. The in-surface curvature drift frequency $\omega_{\kappa}^\alpha$ arises from the curvature of field lines within the flux surface. The radial curvature drift frequency $\omega_{\kappa}^q$ is similar to $\omega_{\kappa}^\alpha$ but is projected radially. The total magnetic drift frequency ${\omega}_{M N}$ combines curvature and grad-B drifts, accounting for the full magnetic drift vector in the field-line-following geometry. The Shafranov shift $\alpha_{\mathrm{MHD} }$ captures the outward displacement of flux surfaces due to plasma pressure. The higher the pressure gradient, the larger the shift. This impacts local magnetic shear and can alter ballooning mode stability. The local magnetic shear $|s_{\mathrm{local}}|$ measures how field line pitch changes with radius at a given poloidal location (hence the `local' shear) — or how quickly neighboring field lines shear past each other. High magnetic shear stretches and tears turbulent eddies, often suppressing turbulence. But near-zero shear can allow zonal flows and enhance transport; $|s_{\mathrm{local}}|$ (along with $\alpha_\mathrm{MHD}$) is particularly important for ballooning stability \cite{Greene1981}.

\begin{figure}[bt]
    \centering
    \begin{subfigure}[t]{0.49\textwidth}
    \centering
    \includegraphics[width=1.0\textwidth]{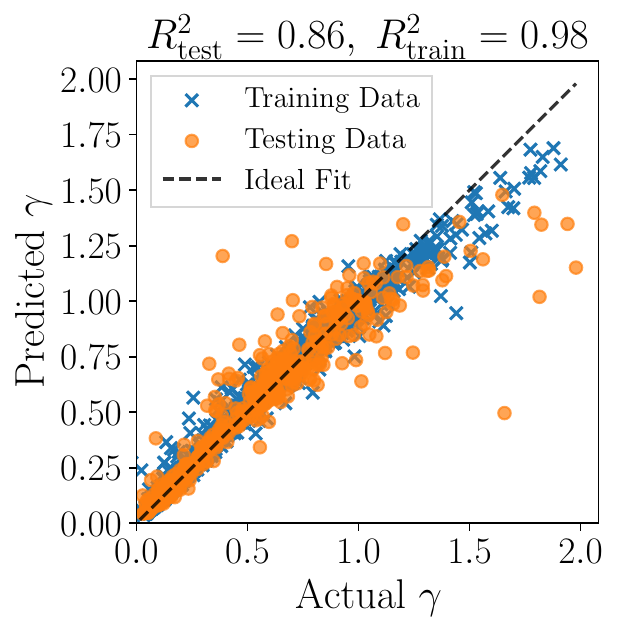}
    \end{subfigure}
    \caption{Random Forest model results for KBM growth rates in NSTX 132543 across a range of binormal wavenumbers. All quantities with a poloidal dependence are evaluated at the outboard midplane.}
    \label{fig:RF_KBM_model_NSTX}
\end{figure}

\begin{figure*}[bt]
    \centering
    \begin{subfigure}[t]{0.49\textwidth}
    \centering
    \includegraphics[width=1.0\textwidth]{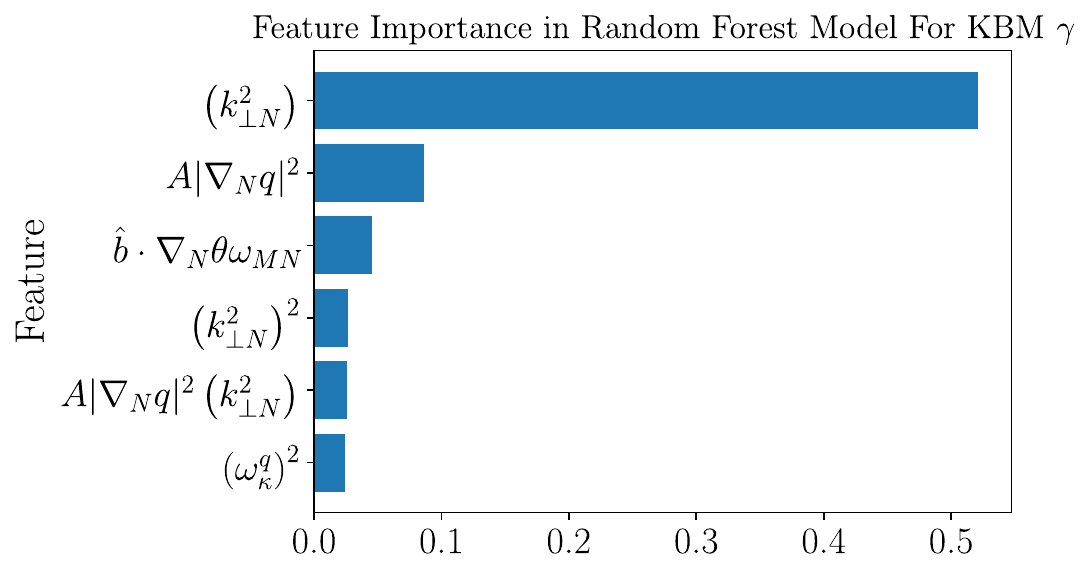}
    \caption{}
    \end{subfigure}
    \centering
    \begin{subfigure}[t]{0.49\textwidth}
    \centering
    \includegraphics[width=1.0\textwidth]{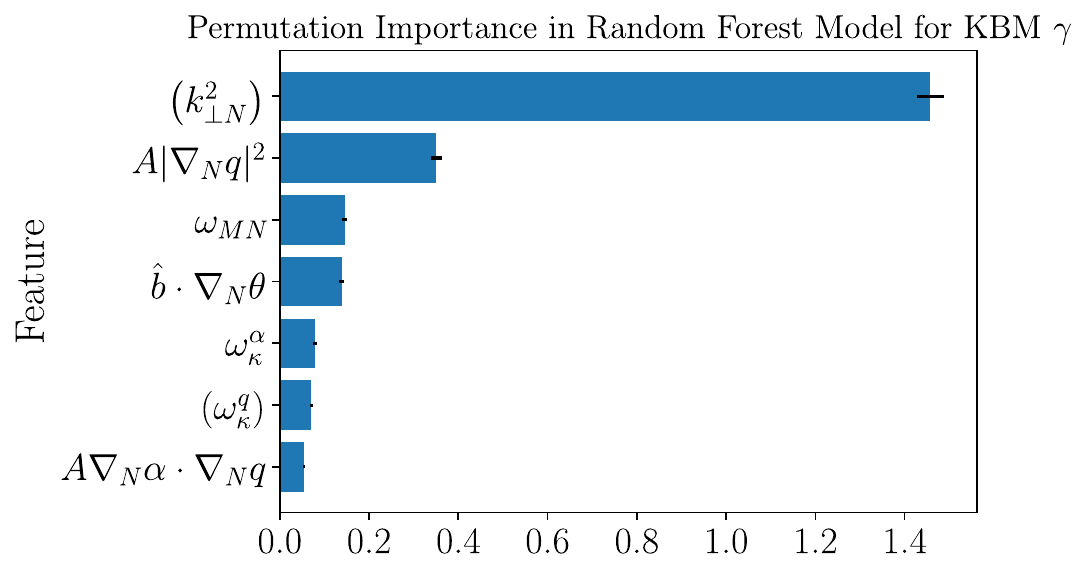}
    \caption{}
    \end{subfigure}
    \centering
    \begin{subfigure}[t]{0.49\textwidth}
    \centering
    \includegraphics[width=1.0\textwidth]{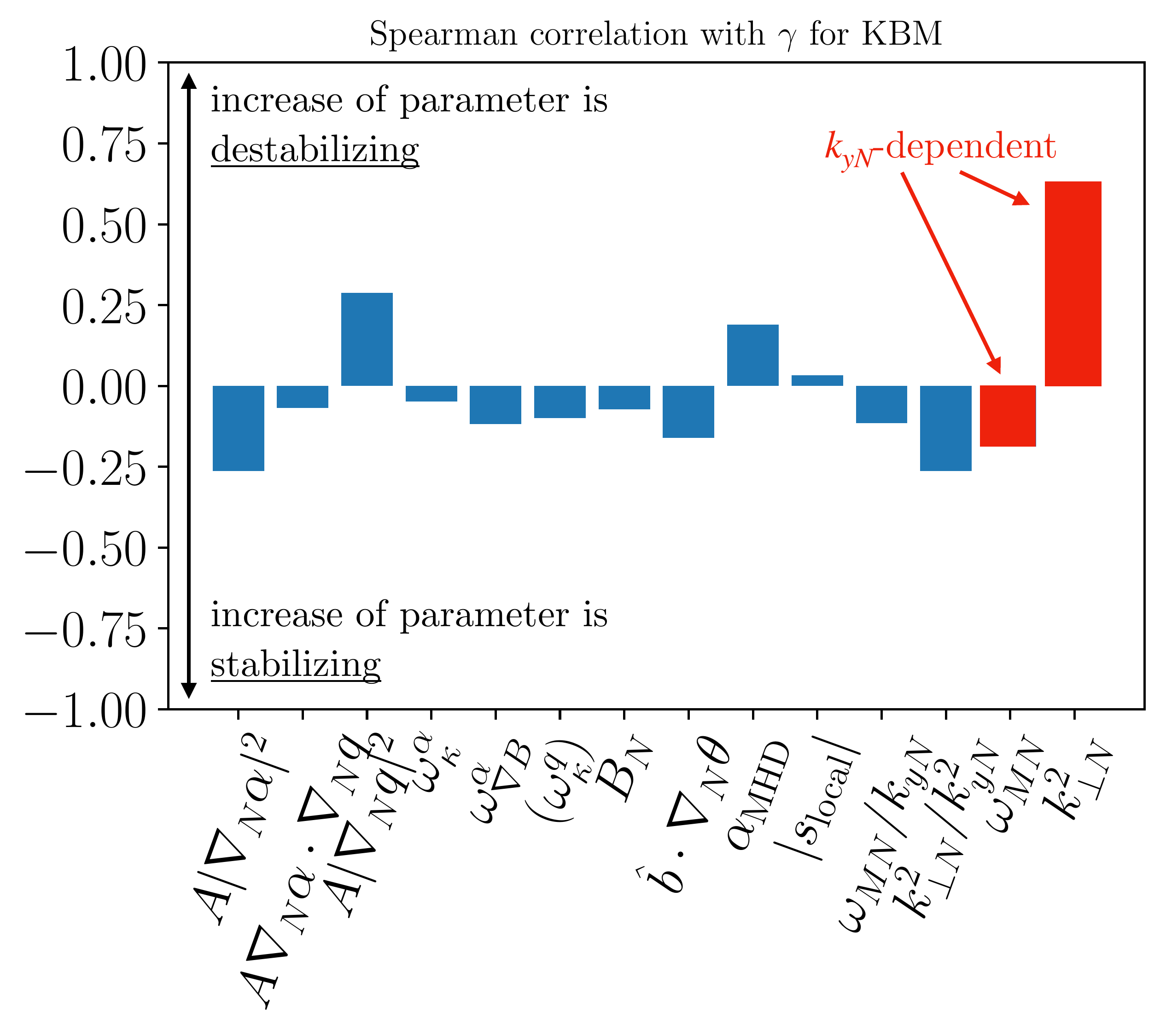}
    \caption{}
    \end{subfigure}
    \centering
    \begin{subfigure}[t]{0.49\textwidth}
    \centering
    \includegraphics[width=1.0\textwidth]{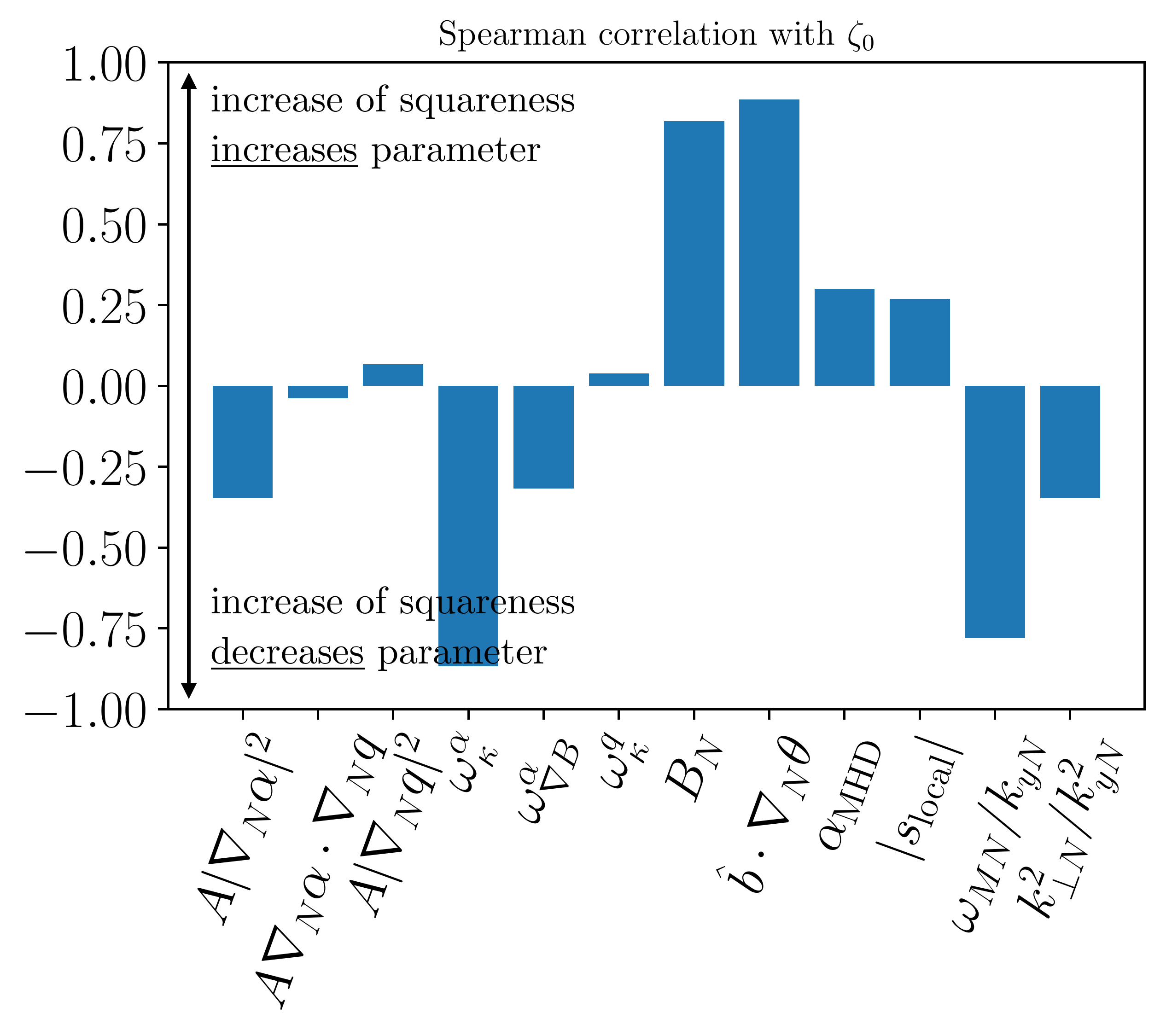}
    \caption{}
    \end{subfigure}
    \caption{Random Forest model results for KBM growth rates in NSTX 132543 across a range of binormal wavenumbers. All quantities with a poloidal dependence are evaluated at the outboard midplane. (a) Feature importance for KBM $\gamma$ model, (b) Permutation importance for KBM $\gamma$ model, (c) Spearman correlation of features with $\gamma$, (d) Spearman correlation of features with $\zeta_0$ (we exclude $k_y$-dependent quantities $k^2_{\perp N}$ and ${\omega}_{MN} $ because $k_{yN}$ is not a property of an equilibrium).}
    \label{fig:RF_KBM_model_NSTX_stats}
\end{figure*}

\subsection{Model for Growth Rates} \label{sec:RF_growth_rates}

Before training a RF model on the gyrokinetic growth rates in NSTX, we prepare the filtered training set. Since we only focus on KBM stability, all modes that are not KBM are filtered out. Additionally, because we are interested in possible ELM-free discharges, we only consider KBMs that are at pedestal gradients lower than those along the KBM first stability boundary. We then impose two further filter conditions: (i) we only consider features with an absolute Spearman correlation value less than 0.7 -- features that are too strongly correlated with others are removed (see \Cref{fig:spearman_NSTX} for the reduced Spearman matrix and \Cref{fig:spearman_NSTX_full} in \Cref{app:spearman_corr_matrix} for the Spearman correlation matrix of all the features), (ii) datapoints with any feature that is more than 5 medians away from the median are removed. After cleaning the data, we randomly separate it into two groups: training and test data, constituting 90\% and 10\% of the total cleaned dataset. The training time was less than one second on a single AMD Opteron 3.1 GHz CPU.

The filtered database of gyrokinetic simulations is input to a RF model to calculate the linear growth rates $\gamma$ for KBMs with the features in \Cref{eq:OMP_features}. A commonly used figure of merit for the accuracy of the RF model is the coefficient of determination $R^2$, defined as
\begin{equation}
R^2 = 1 - \frac{SS_{\mathrm{res}}}{SS_{\mathrm{tot}}}, 
\end{equation}
where $SS_{\mathrm{res}}$ is the residual sum of squares and $SS_{\mathrm{tot}}$ is the total sum of squares.

The model employs a machine learning pipeline consisting of three steps: 
(1) standardization of input features to zero mean and unit variance, 
(2) expansion of the feature space through second-degree polynomial feature generation, 
and (3) non-linear regression using a random forest regressor with a fixed random state for reproducibility. 

Shown in \Cref{fig:RF_KBM_model_NSTX}, $R^2$ of the training data has $R^2_{\mathrm{train} } = 0.98$ and on the test data, $R^2_{\mathrm{test} } = 0.86$. To test the relative importance of each term in our RF model, we plot the feature importance in \Cref{fig:RF_KBM_model_NSTX_stats}(a), which shows that the wavenumber-dependent $k_{\perp N}^2$ is by far the most important for predicting the KBM $\gamma$. Note that composite terms such as $\hat{\mathbf{ b} } \cdot \nabla_B \theta \omega_{MN}$ can appear because the RF model is quadratic. In \Cref{fig:RF_KBM_model_NSTX_stats}(b) we plot the permutation importance, which shows the importance of each feature for accurately predicting the KBM growth rate $\gamma$ in the RF model. Permutation importance is calculated by permuting the input features and calculating the resulting decrease in model accuracy. Features with higher permutation importance cause larger prediction errors when permuted. \Cref{fig:RF_KBM_model_NSTX_stats}(b) shows again that $k_{\perp N}^2$ is by far the most important for predicting $\gamma$ for the KBM, followed by $A |\nabla_N q|^2$.%

In \Cref{fig:RF_KBM_model_NSTX_stats}(c), we plot the Spearman correlation of all the features with the KBM growth rate $\gamma$. The most important features for feature importance and permutation importance -- $k_{\perp N}^2$ and $A |\nabla_N q|^2$ -- are most strongly correlated with $\gamma$.

Because we find that higher plasma squareness $\zeta_0$ gives higher growth rates for KBM near first stability, it is useful to calculate how $\zeta_0$ correlates with the input features. In \Cref{fig:RF_KBM_model_NSTX_stats}(d), we plot the Spearman correlation of all the features with the plasma squareness $\zeta_0$ across all of the flux surfaces and different equilibria considered (we remove $k_{yN}$-dependent quantities $k^2_{\perp N}$ and ${\omega}_{MN}$ because $k_{yN}$ is not a property of an equilibrium). The two Figures can be used in tandem to understand the effect of squareness through each term. The four features (independent of $k_{yN}$) that are correlated most strongly with the KBM growth rate -- $A|\nabla_N \alpha|^2$, $A|\nabla_q \alpha|^2$, $\alpha_\mathrm{MHD}$, and $k_{\perp N}^2/k_{yN}^2$ -- all change with higher squareness (\Cref{fig:RF_KBM_model_NSTX_stats}(d)) in such a way that increases the growth rate. We cannot correlate $k_{\perp N}^2$ or $\omega_{MN}$ with plasma squareness because both quantities depend on $k_{yN}$, which is not a feature of the equilibrium. Because of this limitation, we also calculate correlations with $k_{\perp N}^2/ k_{yN}^2$ and $\omega_{MN}/k_{yN}$, both of which are independent of $k_{yN}$. Physically, higher $k_{\perp N}^2/ k_{yN}^2$ \cite{Parisi2020,Parisi2022} corresponds to stronger FLR damping at a given $k_{yN}$ value. Hence, higher squareness has weaker mode damping effects (smaller $k^2_{\perp N}/k_{yN}$) while also driving stronger instability through higher $\alpha_\mathrm{MHD}$. This tends to destabilize KBM, which gives pedestals with lower average gradients.

\subsection{Model for Proximity to KBM Stability Boundary}

\begin{figure}[!tb]
    \centering
    \begin{subfigure}[t]{0.45\textwidth}
    \centering
    \includegraphics[width=1.0\textwidth]{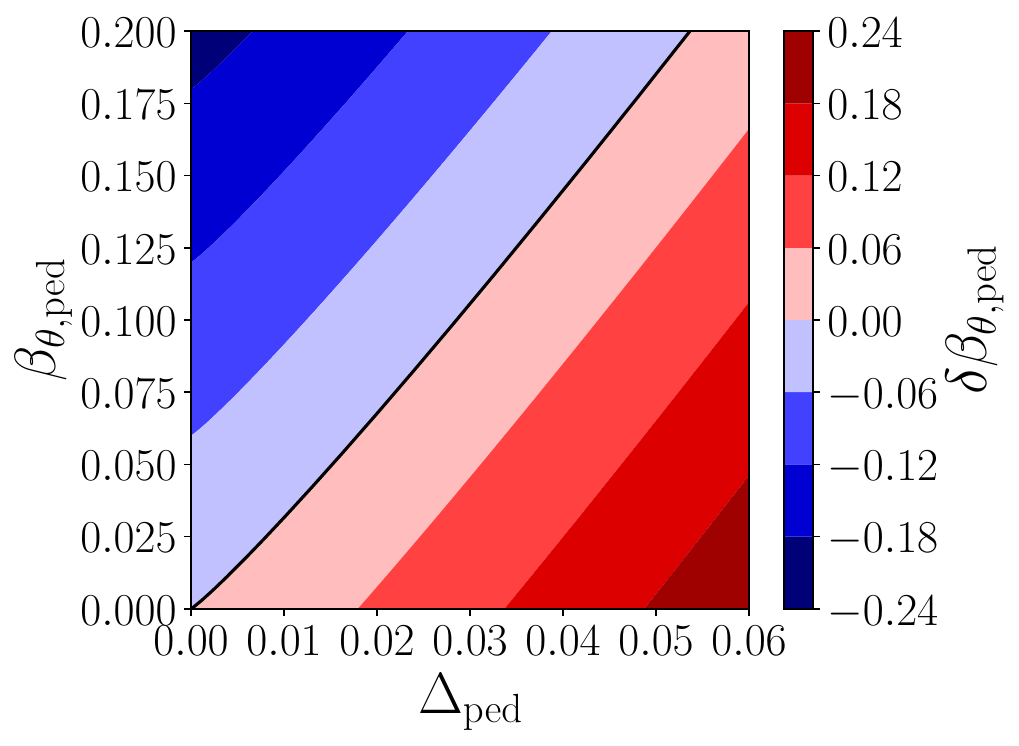}
    \end{subfigure}
    \caption{$\delta \beta_{\theta,\mathrm{ped} }$ for MAST-U 48339 with $\zeta_0=-0.12$.}
    \label{fig:model2}
\end{figure}

\begin{figure}[!tb]
    \begin{subfigure}[t]{0.48\textwidth}
    \centering
    \includegraphics[width=1.0\textwidth]{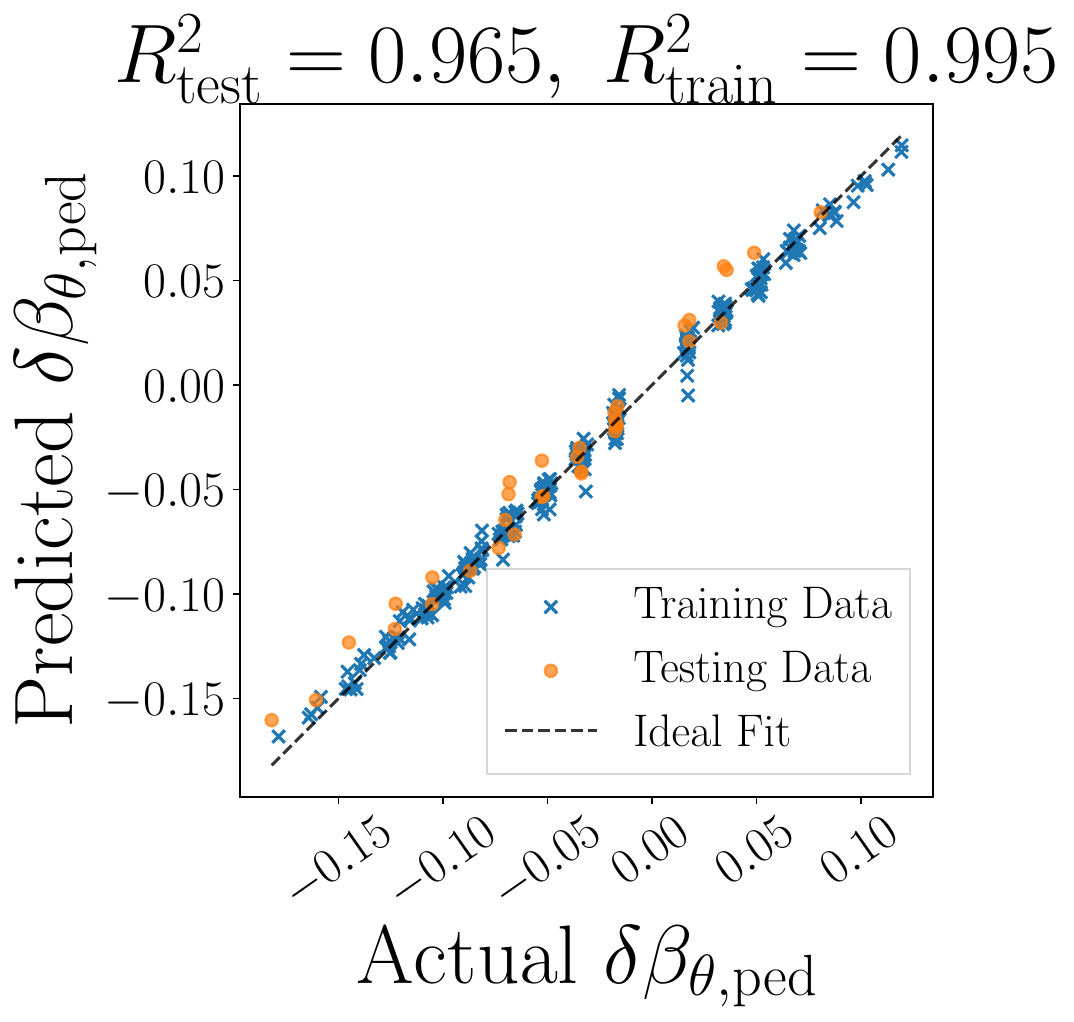}
    \caption{}
    \end{subfigure}
    \centering
    \begin{subfigure}[t]{0.48\textwidth}
    \centering
    \includegraphics[width=1.0\textwidth]{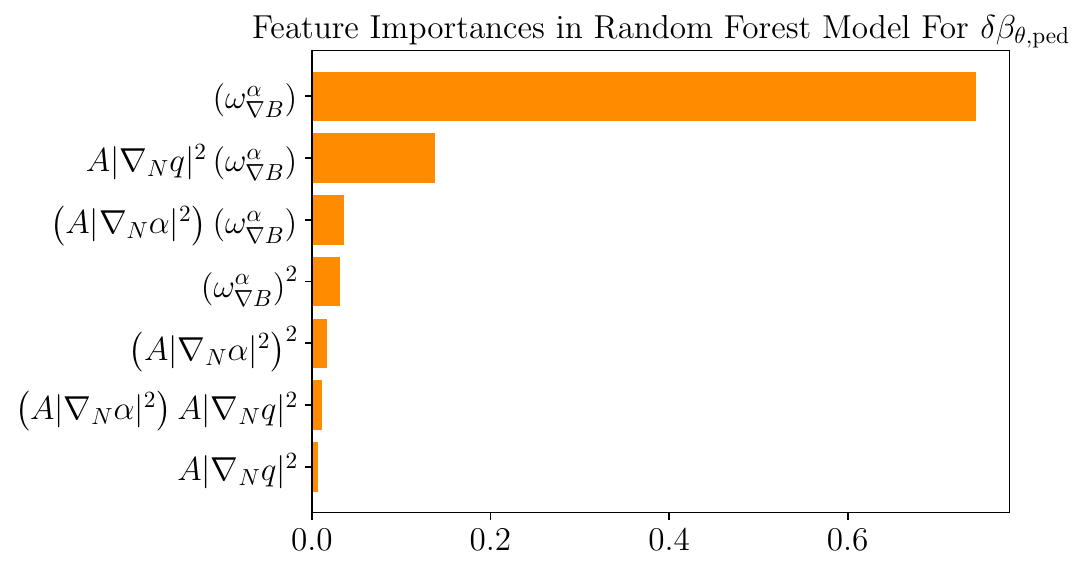}
    \caption{}
    \end{subfigure}
     ~
    \begin{subfigure}[t]{0.48\textwidth}
    \centering
    \includegraphics[width=1.0\textwidth]{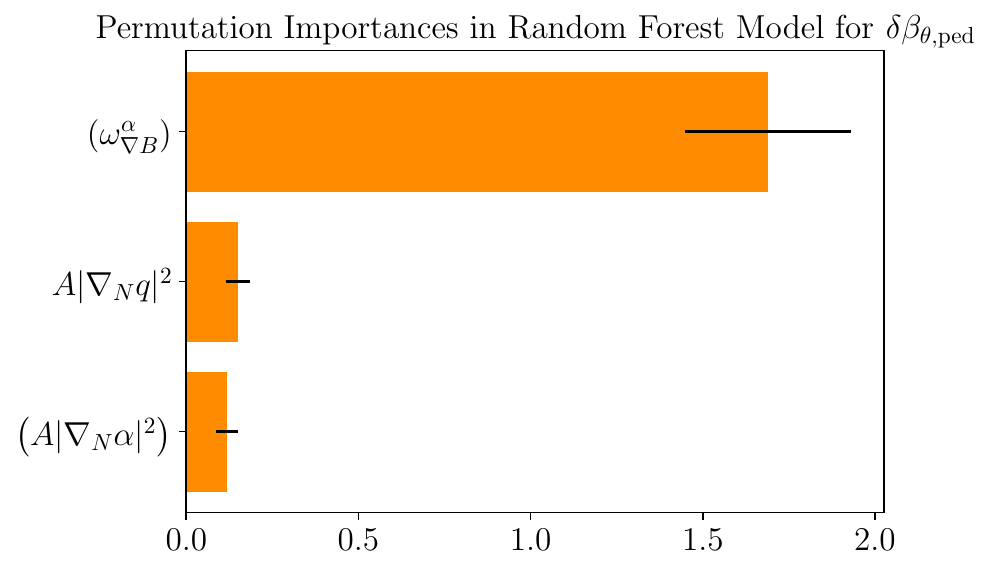}
    \caption{}
    \end{subfigure}
    \caption{Random Forest model results for $\delta \beta_{\theta,\mathrm{ped} }$ in MAST-U 48339. All quantities with a poloidal dependence are evaluated at the outboard midplane. (a) Predicted $\delta \beta_{\theta,\mathrm{ped} }$ versus actual $\delta \beta_{\theta,\mathrm{ped} }$, (b) Feature importance for $\delta \beta_{\theta,\mathrm{ped} }$ model, (c) Permutation importance for $\delta \beta_{\theta,\mathrm{ped} }$ model.}
    \label{fig:model_mastu}
\end{figure}
We now calculate proximity to the first stability boundary for MAST-U using a RF model and then evaluate proximity to the first stability boundary across squareness and triangularity. 

We define the distance from the first stability boundary at fixed pedestal width as
\begin{equation}
\delta \beta_{\theta, \mathrm{ped} } \equiv \beta_{\theta, \mathrm{ped, 1st} } - \beta_{\theta, \mathrm{ped} }.
\end{equation}
Larger positive values of $\delta \beta_{\theta, \mathrm{ped} }$ indicate that the equilibrium is farther below the first stability boundary, and larger negative values predict the equilibrium is farther above the first stability boundary. To obtain discharges farther from the ELM boundary, we generally desire larger $\delta \beta_{\theta, \mathrm{ped} }$. An example of $\delta \beta_{\theta, \mathrm{ped} }$ is shown for a MAST-U equilibrium in \Cref{fig:model2}(a): pedestals with $\delta \beta_{\theta, \mathrm{ped} } > 0$ are below the first KBM stability boundary. As before, the input features for the RF model evaluate geometric coefficients at the OMP. However, unlike the previous RF model for the KBM growth rate where seven pedestal flux surfaces were considered per equilibrium, here we only consider the pedestal mid-radius location. This is because we wish to predict $\delta \beta_{\theta, \mathrm{ped} }$ (which is a property of the equilibrium), whereas the RF model in the previous section predicted $\gamma$ (which is a property of the flux surface and wavenumber). The RF model is trained on $\delta \beta_{\theta, \mathrm{ped} }$ as a function of the geometric parameters.

\begin{figure*}[tb]
    \centering
    \includegraphics[width=0.68\textwidth]{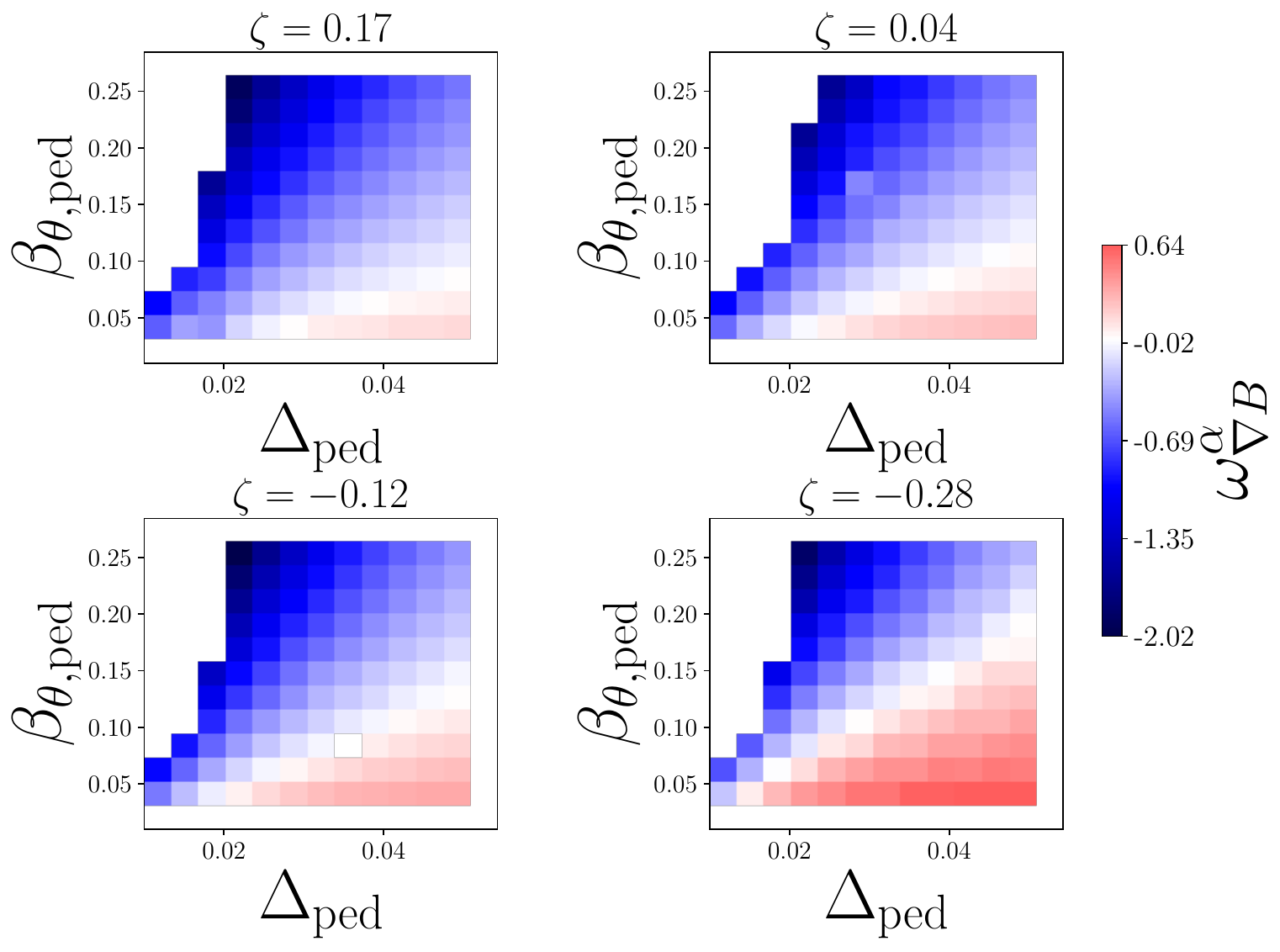}
    \caption{Values of $\omega_{\nabla B}^{\alpha}$ across pedestal width and height for four MAST-U squareness values.}
    \label{fig:model2_gradBlocal}
\end{figure*}

\begin{figure}[b]
    \centering
    \begin{subfigure}[t]{0.45\textwidth}
    \centering
    \includegraphics[width=1.0\textwidth]{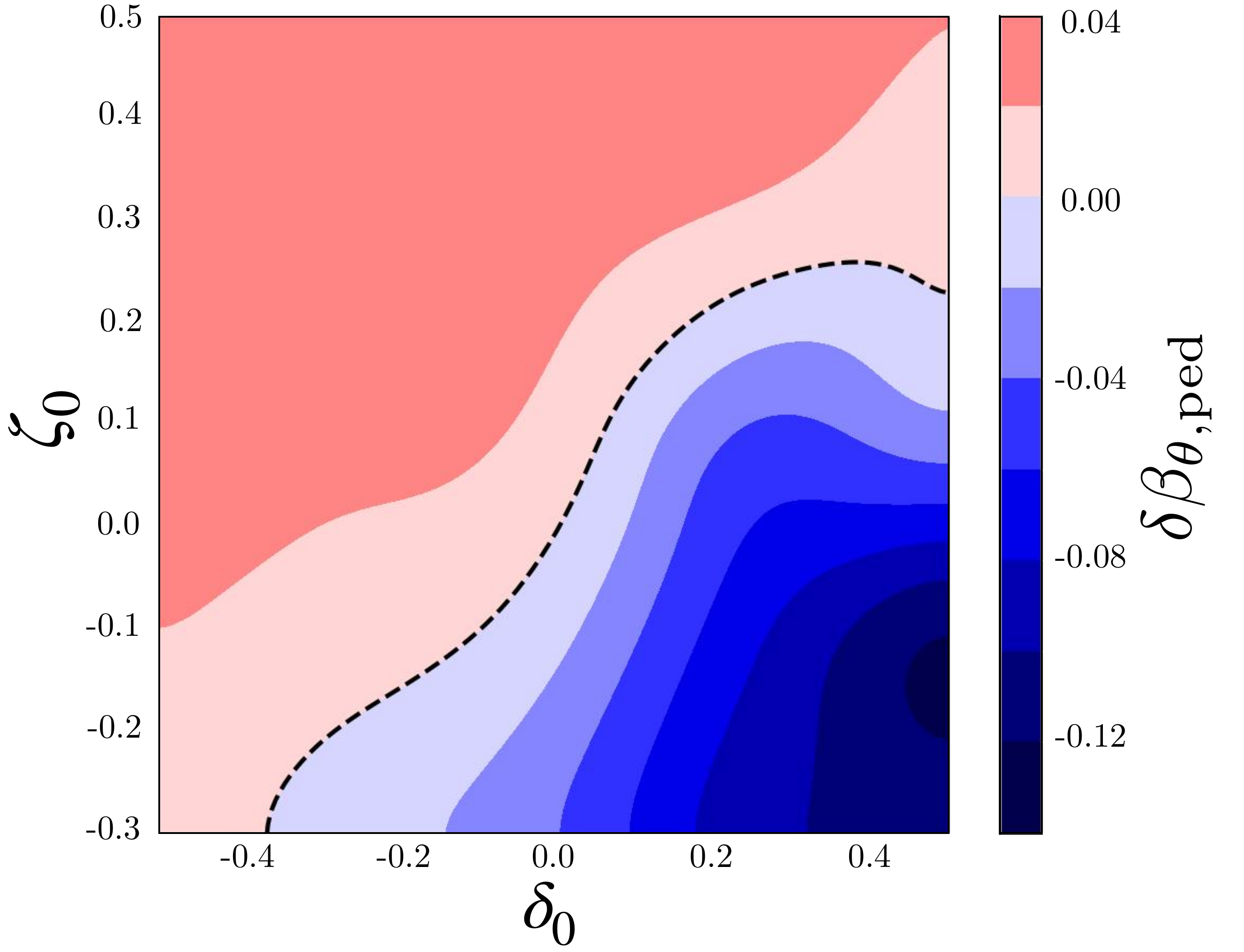}
    \end{subfigure}
    \caption{Distance to the first stability boundary for equilibria MAST-U equilibria based on 48339 with varying squareness $\zeta_0$ and triangularity $\delta_0$.}
    \label{fig:model2_second}
\end{figure}

Shown in \Cref{fig:model_mastu}(a), $R^2$ of the training data has $R^2_{\mathrm{train} } = 0.995$ and on the test data, $R^2_{\mathrm{test} } = 0.965$. This is a remarkably accurate model for the pedestal height, although extensive testing and training on a wider dataset is needed before such a model could be deployed in the same manner as previous machine learning pedestal models such as EPEDNN \cite{Meneghini_2017} and other recent works \cite{Zeger2021,Gillgren2022,Gillgren2023,Jarvinen2024,Joung2024,Parisi2025c}. Shown in \Cref{fig:model_mastu}(b) and (c), by far the most important parameter for predicting $\delta \beta_{\theta, \mathrm{ped} }$ in the RF model is the in-surface grad-B magnetic drift $\omega_{\nabla B}^{\alpha}$. It is important to note that $\omega_{\nabla B}^{\alpha}$ is strongly correlated with several other features that we removed in our RF model: $\alpha_{\mathrm{MHD} }$, $|s_{\mathrm{local} }|$, and $\hat{s}$. Therefore, we are currently unable to causally distinguish between the effect of $\omega_{\nabla B}^{\alpha}$, $\alpha_{\mathrm{MHD} }$, $|s_{\mathrm{local} }|$, and $\hat{s}$ on $\delta \beta_{\theta, \mathrm{ped} }$. We plot $\omega_{\nabla B}^{\alpha}$ versus $\beta_{\theta, \mathrm{ped}}$ and $\Delta_{\mathrm{ped}}$ in \Cref{fig:model2_gradBlocal} for the four squareness values in MAST-U. Higher plasma squareness is strongly correlated with more negative $\omega_{\nabla B}^{\alpha}$ values, and one can see the similarity to \Cref{fig:squareness_first_second_stab_MASTU} of the shifting KBM 1st stability boundary.

We now perform a numerical experiment to calculate how $\delta \beta_{\theta, \mathrm{ped} }$ varies with squareness and triangularity. We generate a new set of equilibria at fixed $\beta_{\theta, \mathrm{ped}}$ and $\Delta_{\mathrm{ped}}$ with varied triangularity and squareness starting from the MAST-U equilibrium. By evaluating the geometric coefficients in each of the new equilibria, the RF model predicts $\beta_{\theta, \mathrm{ped}}$, shown in \Cref{fig:model2_second}. Equilibria with higher positive squareness and more negative triangularity are predicted to have larger $\delta \beta_{\theta, \mathrm{ped} }$ values. Such pedestals would have lower average gradients and may be more likely to be ELM-free. However -- we must be cautious, especially with varying triangularity, because of how changing the triangularity also changes the ELM stability boundary.

\section{Summary} \label{sec:summary}

Using gyrokinetic and peeling-ballooning-mode simulations of NSTX and MAST-U pedestals, we predict that sufficiently high plasma squareness will give rise to ELM-free pedestals. This is caused by peeling-ballooning-mode stability being relatively unaffected by plasma squareness, but gyrokinetic stability degrading. As has been suggested before \cite{Snyder2009,Osborne2015}, by deliberately degrading gyrokinetic stability, the pedestal gradients decrease, moving the pedestal further from the ELM unstable region.

At low aspect ratio, the first stability branch of kinetic-ballooning-mode stability can support steeper gradients than at higher aspect ratio \cite{Parisi_2024b}. This might allow STs with high plasma squareness and degraded edge gradients to still remain in H-mode, whereas analogous squareness values might not sustain an H-mode at higher aspect ratio \cite{Nelson2022,Nelson2024b,Nelson2024c}. Furthermore, because the robustness of ELM-free H-modes has not yet been demonstrated in burning plasmas, high-squareness designs may offer more operational flexibility. If H-mode operation at moderate-to-high squareness still features unacceptably violent ELMs, it may still be possible to operate in L-mode at higher squareness while maintaining adequate confinement \cite{Joiner2010}, similar to negative triangularity L-mode \cite{Sauter2014,Austin2019,Nelson2024b,Wilson2024}.

We trained a random forest machine learning model on the geometric coefficients appearing in the gyrokinetic equation in order to predict the KBM growth rate and the proximity to the KBM pedestal width-height scaling for different equilibrium shaping. Similar approaches have been used recently in stellarator geometries to better understand the effects of shaping on plasma stability and turbulence \cite{Stroteich2022seeking,Roberg2023,Landreman2025doesiontemperaturegradient}.

\section{Acknowledgements}

This work was supported by the U.S. Department of Energy under contract numbers DE-AC02-09CH11466, DE-SC0022270, DE-SC0022272. The United States Government retains a non-exclusive, paid-up, irrevocable, world-wide license to publish or reproduce the published form of this manuscript, or allow others to do so, for United States Government purposes.

\section{Code and Data Availability.}

Part of the data analysis was performed using the OMFIT integrated modeling framework \cite{OMFIT2015}, Pyrokinetics \cite{Patel2024}, and the Github projects \texttt{gk\_ped} \cite{Parisi2023a}, \texttt{ideal-ballooning-solver} \cite{Gaur2023a}. The data that support the findings of this study will be made openly available on an online repository upon publication.

\appendix

\section{Pedestal Parameters} \label{sec:peddefinitions}

In this appendix, we describe the main pedestal parameters used in the paper. The pedestal profiles are parameterized \cite{Snyder2009b,Parisi_2024} as,
\begin{equation}
\begin{aligned}
& n_e(\psi_N) = n_{e,\mathrm{core}} \mathrm{H} \left[  \psi_{\mathrm{ped,n_e}}  -  \psi_N \right] (1-\psi_N^{\alpha_{n_1}})^{\alpha_{n_2}} + \\
& A_n \bigg{[} n_{e0} \left( t_1 - \tanh \left(\frac{\psi_N-\psi_{\mathrm{mid,n_e}}}{ S_{\Delta} \Delta_{n_e} /2} \right) \right) \bigg{]} + n_{\mathrm{e,off}},
\label{eq:1}
\end{aligned}
\end{equation}
\begin{equation}
\begin{aligned}
& T_e(\psi_N) = T_{e,\mathrm{core}} \mathrm{H} \left[  \psi_{\mathrm{ped,T_e}} -  \psi_N \right] (1-\psi_N^{\alpha_{T_1}})^{\alpha_{T_2}} +  \\
& A_T \bigg{[} T_{e0} \left( t_1 - \tanh \left(\frac{\psi_N-\psi_{\mathrm{mid,T_e}}}{ S_{\Delta} \Delta_{T_e}/2} \right) \right) \bigg{]} + T_{\mathrm{e,off}}.
\label{eq:2}
\end{aligned}
\end{equation}
Here, $H$ is a step function, $\psi_N$ is the normalized poloidal flux (0 at the magnetic axis and 1 at the separatrix), $n_{e,\mathrm{core}}$, $T_{e,\mathrm{core}}$, $n_{e0}$, and $T_{e0}$ are constants, $\Delta_{n_e}$ and $\Delta_{T_e}$ are electron density and temperature pedestal widths, and ${\alpha_{\{n,T,J\},\{1,2\}}}$ are exponents. Scalar quantities $A_{n}$, $A_{T}$, and $S_{\Delta}$ rescale the pedestal density, temperature, and width. The pedestal heights are
\begin{equation}
n_{e,\mathrm{ped}} = n (\psi_{\mathrm{ped,n_e}}), \; T_{e,\mathrm{ped}} = T(\psi_{\mathrm{ped,T_e}}),
\end{equation}
$n_{e,\mathrm{off}}$ and $T_{e,\mathrm{off}}$ are evaluated at $\psi_N = \psi_{\mathrm{ped,n_e}} + \Delta_{n_e}$ and $\psi_N = \psi_{\mathrm{ped,T_e}} + \Delta_{T_e}$ respectively, $t_1 = \tanh \left(1 \right)$, and $\psi_{\mathrm{ped},n_e} = \psi_{\mathrm{mid,n_e}} - \Delta_{n_e} / 2$. The pedestal width is
\begin{equation}
\Delta_{\mathrm{ped}} = (\Delta_{n_e} + \Delta_{T_e})/2,
\end{equation}
the pedestal top location is $\psi_{\mathrm{ped}} = \psi_{\mathrm{mid}} - \Delta_{\mathrm{ped}} / 2$, where $\psi_{\mathrm{mid}} =  (\psi_{\mathrm{mid},n_e}+ \psi_{\mathrm{mid},T_e})/2$. The normalized pedestal height is
\begin{equation}
\beta_{\theta, \mathrm{ped}} = 8 \pi p_{\mathrm{ped}} /\overline{B}_{\mathrm{pol}}^2
\end{equation}
where the pedestal pressure height is $p_{\mathrm{ped}} = 2 p_e (\psi_N = \psi_{\mathrm{ped}} )$, $\overline{B}_{\mathrm{pol}} =4\pi I_p /  l c $ with last-closed-flux-surface circumference $l$ where $I_p$ is the total plasma current and $c$ is the speed of light.

To describe increasing pedestal height with different density and temperature contributions, we use the pedestal pressure rescaling factor $S_p$ to rescale $p_{\mathrm{ped}}$,
\begin{equation}
S_p = \frac{p_{\mathrm{ped,new} }}{p_{\mathrm{ped,orig} }}  = S_T S_n.
\label{eq:Sp}
\end{equation}
Here, $S_T$ and $S_n$ are the temperature and density rescaling factors. We relate $S_T$ and $S_n$ using
\begin{equation}
S_T = \left( S_n \right)^b,
\label{eq:ST_Sn_scaling}
\end{equation}
where $b$ is a real number. For rescaling $p_{\mathrm{ped}}$ we use $S_T, S_n$ according to \Cref{eq:Sp} and $A_T, A_n$ (\Cref{eq:1,eq:2}), which are in general complicated functions, are calculated accordingly.

\section{Geometric definitions} \label{sec:geodefinitions}

\begin{table*}
\caption{Key quantities used in this work: reference quantities, unnormalized quantities, and normalized quantities.}
\begin{ruledtabular}
\centering
  \begin{tabular}{ ccc  }
   Name & Quantity & Definition \\
    \hline
    Gyroradius & $\rho_{r} $ & \\
    Reference length (minor radius) & $a $ & \\
    Reference mass & $m_{r} $ & \\
    Reference temperature & $T_{r} $ & \\
    Reference density & $n_{r} $ & \\
    Reference field & $B_{r} $ & \\
    Reference thermal speed & $v_{tr} $ & $\sqrt{T_r / 2 m_r} $ \\ \hline
    Maxwellian distribution function & $F_{Ms} $ & \\
    Perturbed distribution function & $\delta F_{s} $ & \\
    Non-adiabatic distribution function & $h_{s} $ & $\delta f_s + e \phi F_{Ms} / T_s $ \\
    Fluctuating electrostatic potential & $\phi $ & \\
    Fluctuating parallel potential & $ A_{\parallel} $ & \\
    Fluctuating perpendicular potential & $B_{\parallel} $ & \\
    Equilibrium magnetic field & $\mathbf{B} $ & \\
    Equilibrium magnetic field unit vector & $\hat{\mathbf{ b} } $ & $\mathbf{B}/B $ \\
    Safety factor & $q$ & $(1/2\pi) \int (\mathbf{B} \cdot \nabla \zeta) / (\mathbf{B} \cdot \nabla \theta ) d \theta $ \\
    Magnetic shear & $\hat{s}$ & $(\rho/q)(dq/d\rho)$ \\
    Radial spatial flux-tube coordinate & $x $ & $a q_c / \rho_c \left( \psi_N - \psi_{cN} \right) $ \\
    Minor radius coordinate & $r $ & Flux surface half-diameter \\
    Minor radius & $a $ & Last-closed-flux-surface half-diameter \\
    Normalized flux coordinate & $\rho$ & $r/a$, $\psi_N$, or $\sqrt{\psi_{TN}}$  \\
    Major radius coordinate & $R $ & \\
    Major radius & $R_0 $ & Last-closed-flux-surface major radius \\
    Poloidal fux & $\psi $ & \\
    Toroidal fux & $\psi_T $ & \\
    Binormal spatial coordinate & $y $ & $$ \\
    Poloidal angle & $\theta $ & \\
    Toroidal angle & $\zeta $ & \\
    Perpendicular wavenumber & $\mathbf{ k}_\perp $ & $k_x \nabla x + k_y \nabla y$ \\ \hline
    Radial flux coordinate (normalized) & $\rho_N$ & \\
    Gyroradius (normalized) & $\rho_{*r} $ & $\rho_r / a$ \\
    Non-adiabatic distribution function (normalized) & $h_{Ns}$ & $h_s / (\rho_{*r}  F_{Ms}) $ \\
    Fluctuating electrostatic potential (normalized) & $\phi_{N} $ & $\phi / (\rho_{*r} T_r / e)$ \\
    Fluctuating parallel potential (normalized) & $A_{\parallel N} $ & $\phi / (\rho_{*r} T_r / e)$ \\
    Fluctuating parallel magnetic field (normalized) & $B_{\parallel N} $ & $\phi / (\rho_{*r} T_r / e)$ \\
    Equilibrium magnetic field (normalized) & $B_{N} $ & $B / B_r$ \\
    Temperature (normalized) & $T_{Ns} $ & $T_s / T_r$ \\
    Density (normalized) & $n_{Ns} $ & $n_s / n_r$ \\
    Radial wavenumber (normalized) & $k_{x N} $ & $k_x \rho_r$ \\
    Poloidal wavenumber (normalized) & $k_{y N} $ & $k_y \rho_r$ \\
    Perpendicular wavenumber (normalized) & $k_{\perp N} $ & $k_{\perp} \rho_r$ \\
    Radial spatial flux-tube coordinate (normalized) & $x_N$ & $x / \rho_r$ \\
    Binormal spatial coordinate (normalized) & $y_N$ & $y / \rho_r$ \\
    Poloidal flux coordinate (normalized) & $\psi_N$ & $\psi / (a^2 B_r)$ \\
    Toroidal flux coordinate (normalized) & $\psi_{TN}$ & $\psi_T / (a R_0 B_r)$ \\
    Thermal speed (normalized) & $v_{tNs}$ & $v_{ts} / v_{tr}$ \\
    Parallel velocity coordinate (normalized) & $v_{\parallel N}$ & $v_{\parallel} / v_{tr}$ \\
    Perpendicular velocity coordinate (normalized) & $v_{\perp N}$ & $v_{\perp} / v_{tr}$ \\
    Linearized collision operator (normalized) & $C_{Ns}^l$ & $C_{s}^l / ( \rho_{*r} v_{tr} F_{Ms}/a)$ \\
    Radial curvature magnetic drift frequency (normalized) & $\omega_{\kappa}^q$ & \Cref{eq:curvature_drift_alpha} \\
    Binormal grad-B magnetic drift frequency (normalized) & $\omega_{\nabla B}^\alpha$ & \Cref{eq:nablaBalpha} \\
    Binormal curvature magnetic drift frequency (normalized) & $\omega_{\kappa}^\alpha$ & \Cref{eq:curvature_drift_radial} \\
  \end{tabular}
\end{ruledtabular}
\label{tab:tab0}
\end{table*}

In this appendix, we describe the normalization convention for terms in the normalized gyrokinetic equation. All key gyrokinetic quantities are listed in \Cref{tab:tab0}. To describe the directions perpendicular to the magnetic field, we use the flux coordinates
\begin{equation}
x = \frac{q_c}{r_c B_r} \left( \psi - \psi_c \right), \;\;\; y = \frac{1}{B_r} \frac{\partial \psi}{\partial r} \left( \zeta - q \theta - \nu \right),
\end{equation}
where $q_c$, $\psi_c$ and $r_c$ are the safety factor, poloidal flux divided by $2\pi$, and minor radius evaluated at the flux-tube radial center, $B_r$ is a reference magnetic field typically evaluated on the magnetic axis, $\zeta$ is the toroidal angle, $\theta$ is a poloidal angle, and $\nu (r, \theta)$ is a function that describes the effects of a varying pitch-angle along the magnetic field line \cite{Dhaeseleer1991}.

The gyrokinetic flux-tube code GS2 \cite{Dorland2000} normalizes and non-dimensionalizes all quantities appear in the gyrokinetic system of equations. For example, the dimensionless normalized parallel streaming term is
\begin{equation}
\frac{v_{t,s}}{v_{t,r}} v_{\parallel N} (\hat{\mathbf{ b} } \cdot \nabla_N \theta) \frac{\partial h_{Ns}}{ \partial \theta}.
\end{equation}
Here, a `N' subscript indicates that the quantity is normalized to be dimensionless. For example, velocities are normalized by the thermal velocity of the species $v_{t,s}$, $v_{\parallel N} = v_{\parallel }/ v_{t,s}$, gradients are normalized by the device minor radius $a$, $\nabla_N = a \nabla$, and magnetic fields are normalized by a reference magnetic field $B_N = B/ B_r$ where $B_r$ is typically the field strength on axis. The magnetic drift term is
\begin{equation}
\begin{aligned}
& \mathbf{v}_{MNs} \cdot \mathbf{k}_{\perp N} = \\
& \frac{k_{yN} T_{Ns}}{2 Z_s} \left[ \frac{v_{\perp N}^2 }{2} \left( \omega_{\nabla B}^\alpha + \theta_0 \omega_{\nabla B}^q \right) +  v_{\parallel N}^2 \left( \omega_{\kappa}^\alpha + \theta_0 \omega_{\kappa}^q \right) \right],
\end{aligned}
\end{equation}
where the in-surface $\nabla B$ drift is
\begin{equation}
\omega_{\nabla B}^\alpha = \frac{2}{B_N^2} \frac{d \psi_N}{ d \rho} \left( \hat{\mathbf{ b} } \times \nabla_N B_N \right) \cdot \nabla_N \alpha,
\label{eq:nablaBalpha}
\end{equation}
the in-surface curvature drift is
\begin{equation}
\omega_{\kappa }^\alpha = \omega_{\nabla B}^\alpha + \frac{2}{B_N} \frac{d \psi_N}{ d \rho} \left( \hat{\mathbf{ b} } \times \nabla_N \beta \right) \cdot \nabla_N \alpha,
\label{eq:curvature_drift_alpha}
\end{equation}
and the radial $\nabla B$ and curvature drifts are equal,
\begin{equation}
\omega_{\nabla B}^q = \omega_{\kappa}^q = \frac{2}{B_N^2} \frac{d \psi_N}{ d \rho} \left( \hat{\mathbf{ b} } \times \nabla_N B_N \right) \cdot \nabla_N q.
\label{eq:curvature_drift_radial}
\end{equation}
Here, $d \psi_N / d \rho = (d \psi / d r) (1/a B_a)$ for the poloidal flux $\psi$, minor radial coordinate $r$, and normalized flux function $\rho = r/a$. We use an $\alpha$ superscript to denote a drift in the binormal direction and $q$ for a drift in the radial direction, where $q$ is the safety factor and $\beta = 4\pi p / B^2$, where $p$ is the plasma pressure on the flux surface.

\section{Eigenmode-Averaged RF Model} \label{sec:eigen_average}

In this appendix, we show a RF model with eigenmode-averaged geometric coefficients. The model performs marginally better than simply evaluating the geometric coefficients at the outboard midplane. However, it is less useful because obtaining the KBM eigenmode requires performing gyrokinetic simulations.

We study the effect of eigenmode-averaged geometric coefficients on KBM stability. We define the eigenmode-average for a quantity $g$ as
\begin{equation}
\langle g \rangle_{\phi} \equiv \frac{\int |\phi|^2 g d\theta}{\int |\phi|^2 d\theta},
\end{equation}
where $\phi$ is the electrostatic potential. The datapoints correspond to exactly the same KBMs as in \Cref{sec:RF_growth_rates}.

\begin{figure}[bt]
    \centering
    \begin{subfigure}[t]{0.49\textwidth}
    \centering
    \includegraphics[width=1.0\textwidth]{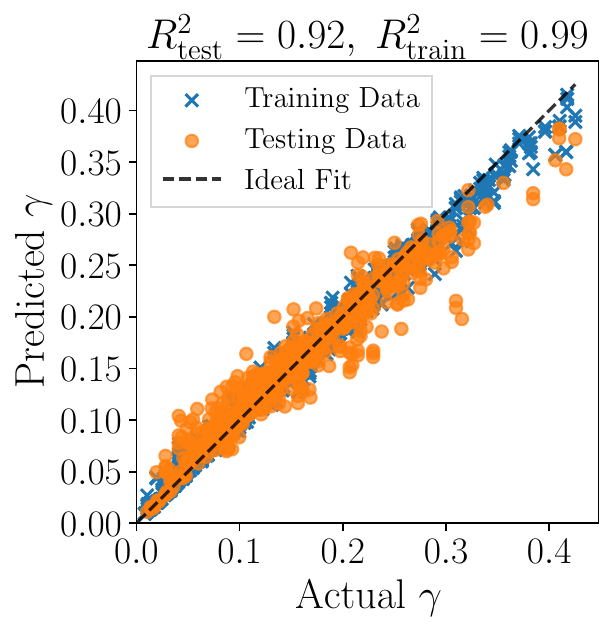}
    \caption{}
    \end{subfigure}
    \caption{Random Forest model results for KBM growth rates in NSTX across a range of binormal wavenumbers. All quantities with a poloidal dependence eigenmode averaged.}
    \label{fig:RF_KBM_model_NSTX_eigenmodeaveraged}
\end{figure}

\begin{figure*}[tb]
    \centering
    \begin{subfigure}[t]{0.49\textwidth}
    \centering
    \includegraphics[width=1.0\textwidth]{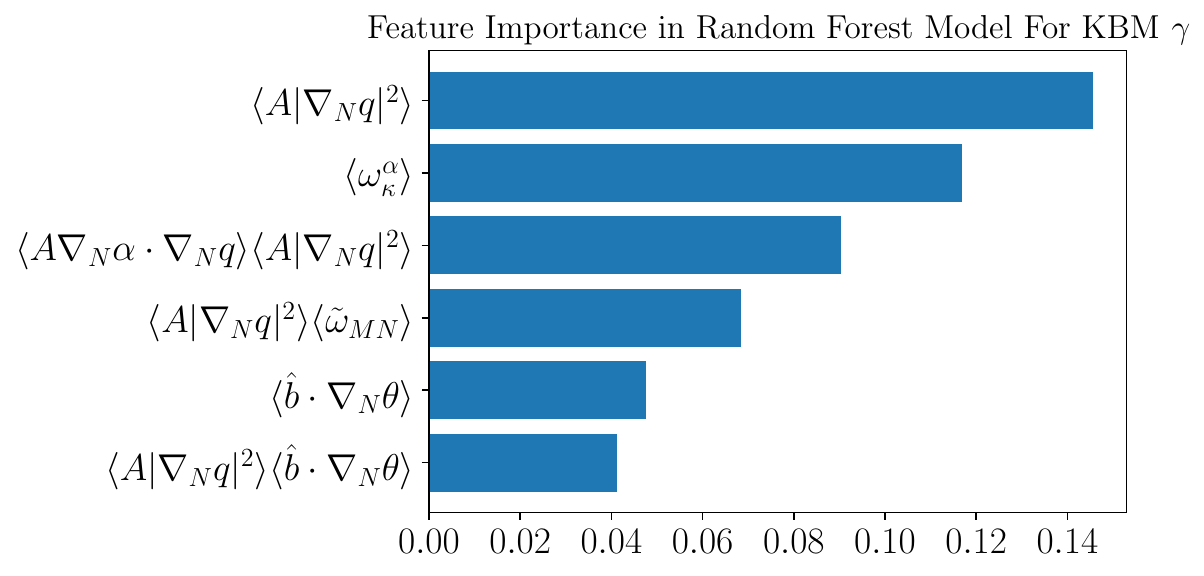}
    \caption{}
    \end{subfigure}
    \centering
    \begin{subfigure}[t]{0.49\textwidth}
    \centering
    \includegraphics[width=1.0\textwidth]{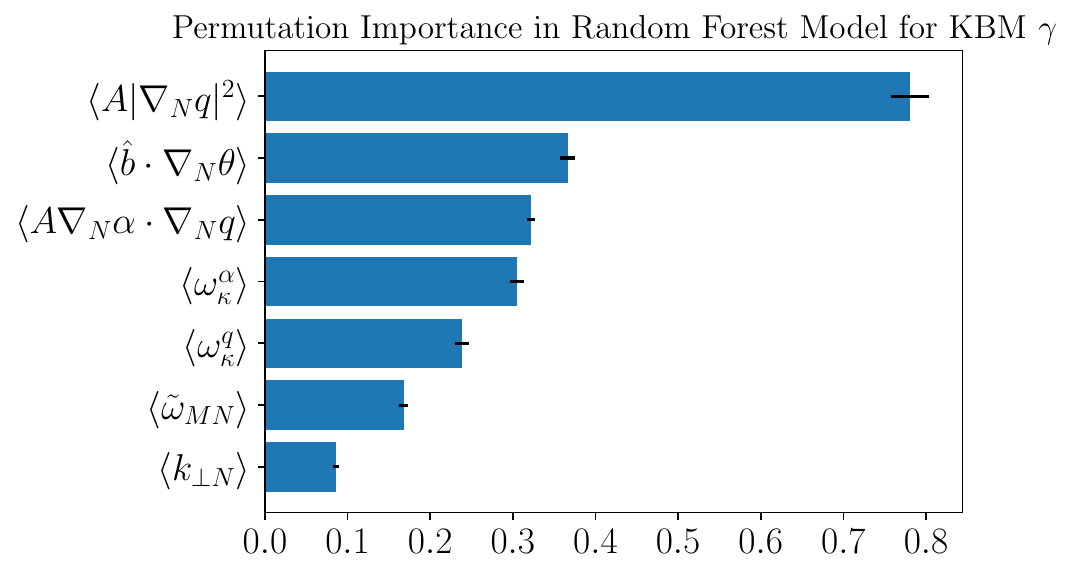}
    \caption{}
    \end{subfigure}
    \centering
    \begin{subfigure}[t]{0.49\textwidth}
    \centering
    \includegraphics[width=1.0\textwidth]{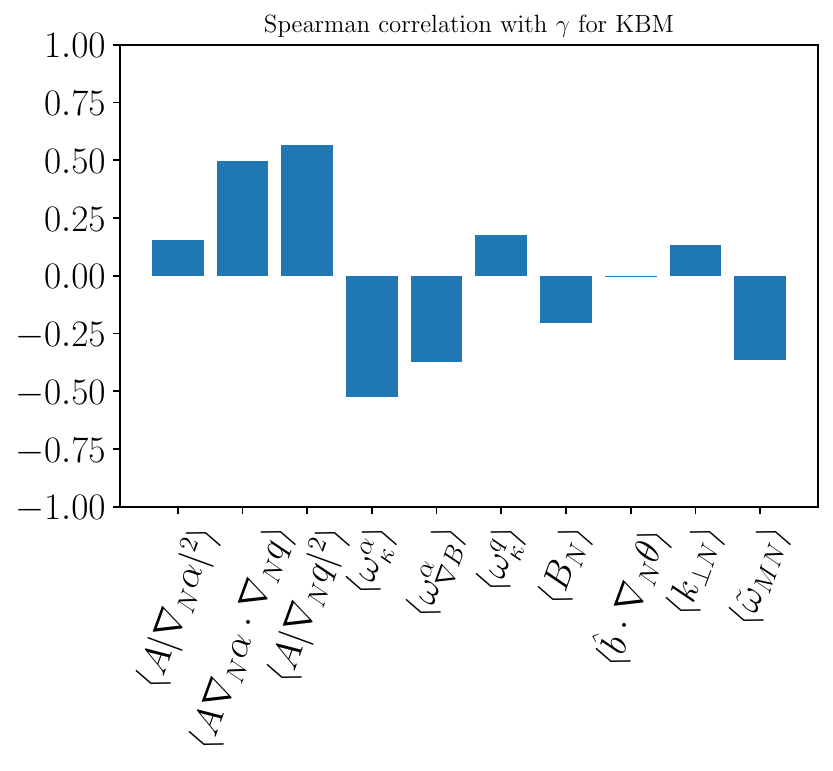}
    \caption{}
    \end{subfigure}
    \centering
    \begin{subfigure}[t]{0.49\textwidth}
    \centering
    \includegraphics[width=1.0\textwidth]{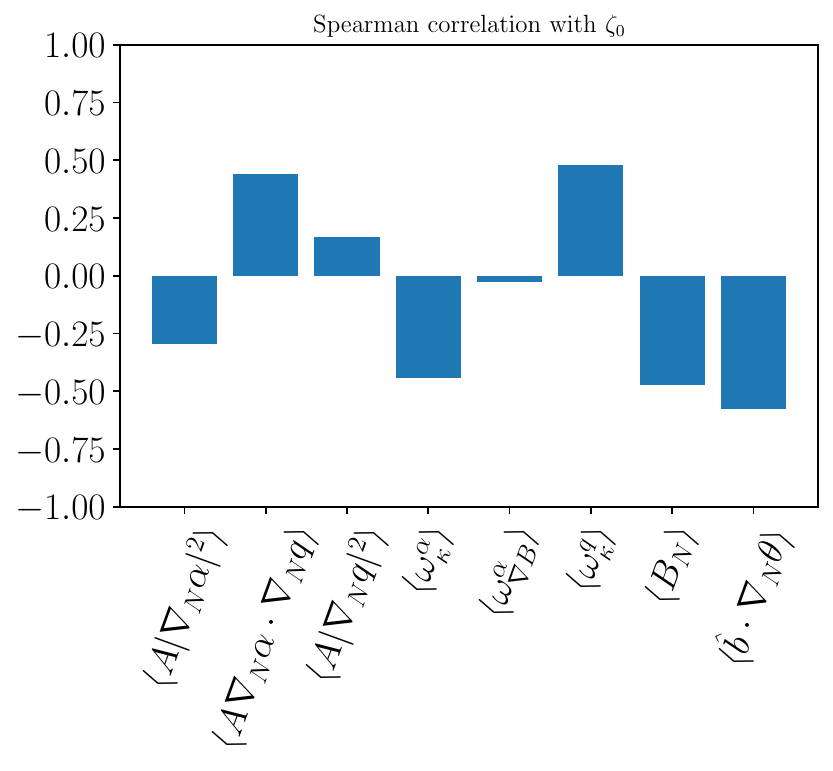}
    \caption{}
    \end{subfigure}
    \caption{Random Forest model results for KBM growth rates in NSTX across a range of binormal wavenumbers. All quantities with a poloidal dependence are eigenmode averaged. (a) Feature importance for KBM $\gamma$ model, (b) Permutation importance for KBM $\gamma$ model, (c) Spearman correlation of features with $\gamma$, (d) Spearman correlation of features with $\zeta_0$ (we exclude $k_y$-dependent quantities $k^2_{\perp N}$ and ${\omega}_{MN} $).}
    \label{fig:RF_KBM_model_NSTX_stats_eigenmodeaveraged}
\end{figure*}

Shown in \Cref{fig:RF_KBM_model_NSTX_eigenmodeaveraged}, $R^2$ of the training data has $R^2_{\mathrm{train} } = 0.99$ and on the test data, $R^2_{\mathrm{test} } = 0.92$, which is marginally better than just using the outboard midplane quantities ($R^2_{\mathrm{train} } = 0.98$ and on the test data, $R^2_{\mathrm{test} } = 0.86$). To test the relative importance of each feature in our RF model, we plot the feature importance in \Cref{fig:RF_KBM_model_NSTX_stats_eigenmodeaveraged}(a), which shows that $\langle A |\nabla_N q|^2 \rangle $ is by far the most important feature for predicting the KBM $\gamma$. In \Cref{fig:RF_KBM_model_NSTX_stats_eigenmodeaveraged}(b) we plot the permutation importance, showing again that $\langle A |\nabla_N q|^2 \rangle $ is by far the most important for predicting $\gamma$. This contrasts with the outboard midplane model, which found the binormal wavenumber-dependent $k^2_{\perp N}$ as the most important feature. The fact that the eigenmode-averaged model finds that explicitly binormal wavenumber-dependent features $k^2_{\perp N}$ and $\langle {\omega}_{MN} \rangle $ are relatively important is likely because the eigenmode-averaging operation contains information about the mode's binormal wavenumer.

\section{Full Spearman Correlation Matrix} \label{app:spearman_corr_matrix}

\begin{figure*}
    \centering
    \begin{subfigure}[t]{0.89\textwidth}
    \centering
    \includegraphics[width=1.0\textwidth]{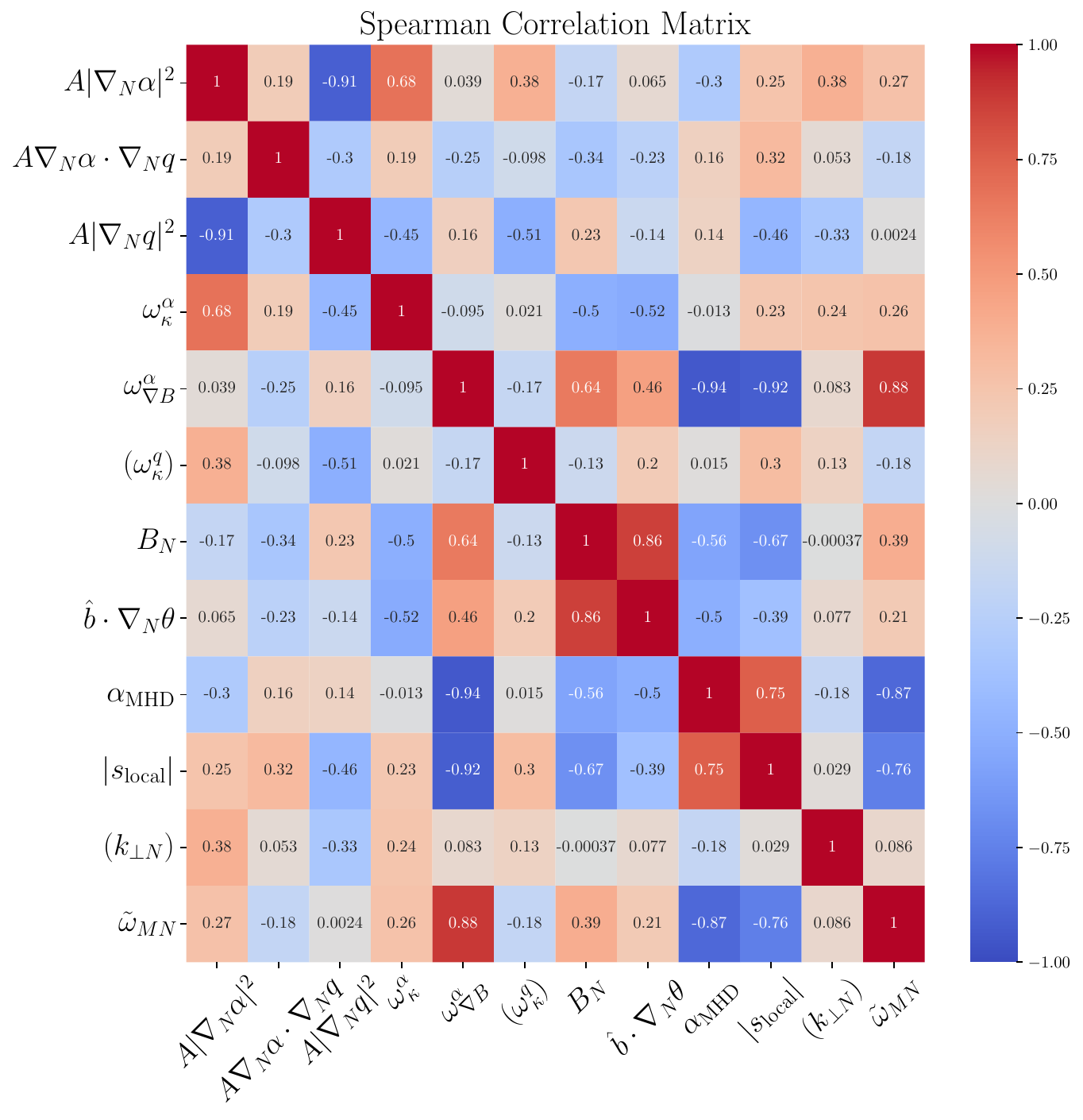}
    \end{subfigure}
    \caption{Spearman correlation of geometric coefficients in NSTX pedestals with six squareness values based on NSTX 132543. Coefficients that have an absolute Spearman value greater than 0.70 are removed from model analysis.}
    \label{fig:spearman_NSTX_full}
\end{figure*}

In this appendix, for completeness we show the full correlation matrix for NSTX 132543 in \Cref{fig:spearman_NSTX_full}. Coefficients that have an absolute Spearman value greater than 0.70 are removed from model analysis and are shown in \Cref{fig:spearman_NSTX}

\bibliographystyle{apsrev4-1} %
\bibliography{EverythingPlasmaBib} %

\end{document}